\documentclass[12pt]{iopart}
\usepackage{iopams}
\usepackage{graphicx}
\usepackage{cite}
\begin{document}

\title{Two-fluid model for heavy electron physics}

\author{Yi-feng Yang$^{1,2}$}

\address{$^1$ Beijing National Laboratory for Condensed Matter Physics, and Institute of Physics, Chinese Academy of Sciences, Beijing 100190, China}
\address{$^2$ Collaborative Innovation Center of Quantum Matter, Beijing 100190, China}
\ead{yifeng@iphy.ac.cn}
\vspace{10pt}
\begin{indented}
\item[]March 2016
\end{indented}

\begin{abstract}
The two-fluid model is a phenomenological description of the gradual change of the itinerant and local characters of the f-electrons with temperature and other tuning parameters and has been quite successful in explaining many unusual and puzzling experimental observations in heavy electron materials. We review some of these results and discuss possible implications of the two-fluid model in understanding the microscopic origin of heavy electron physics.
\end{abstract}

\submitto{\RPP}
%\maketitle

\section{Introduction}

\begin{figure}[t]
\centerline{{\includegraphics[width=0.5\textwidth]{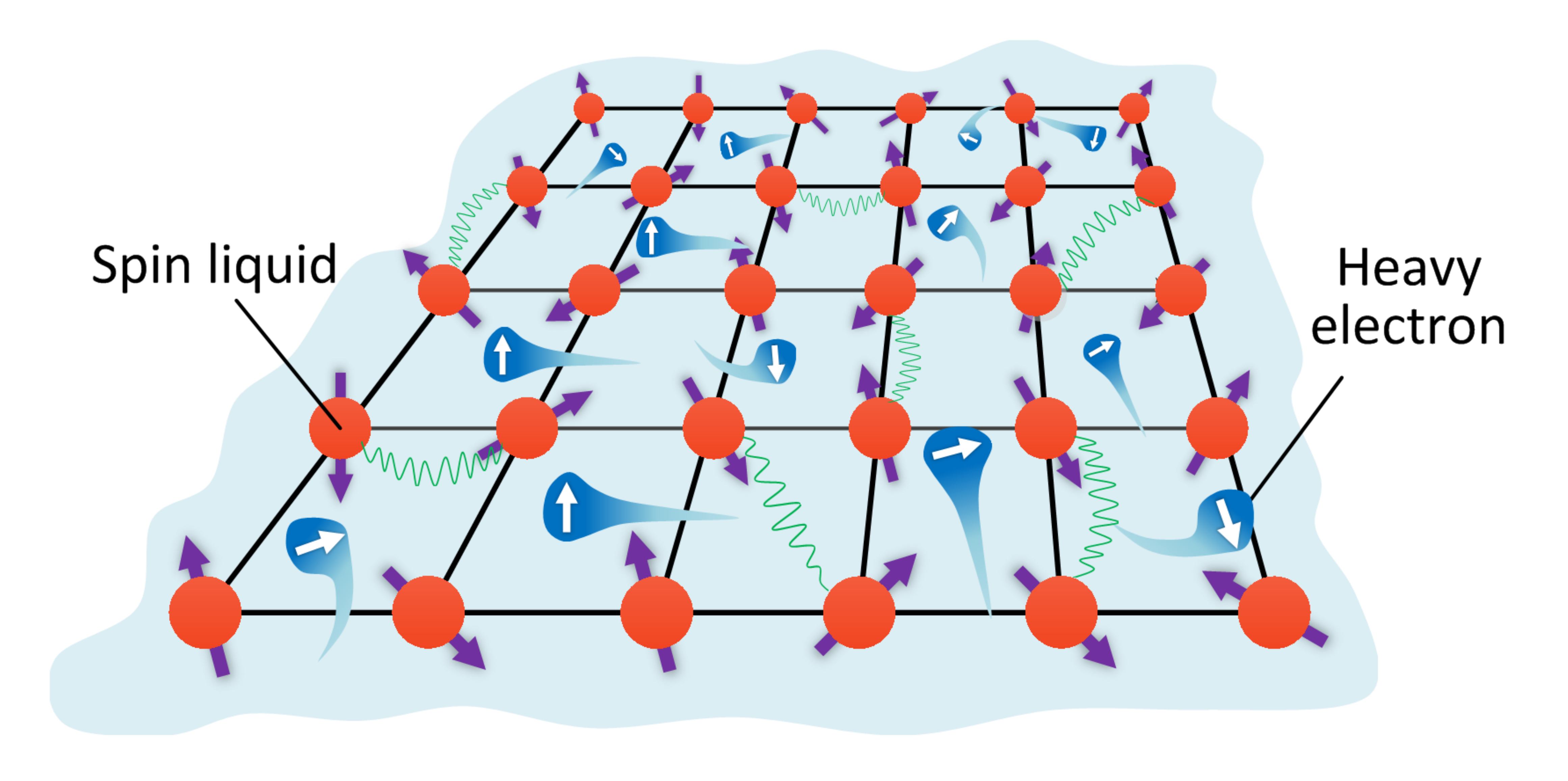}}}
\caption{
{(Color online) Illustration of the two-fluid model in which the antiferromagnetical coupling between conduction electrons and lattice spins gives rise to a renormalized heavy electron fluid (called the Kondo liquid) and a spin liquid of localized moments with reduced strength.}
\label{Fig1}}
\end{figure}

Heavy electron materials are often described as a Kondo lattice that is composed of an array of interacting local moments of 4f or 5f electrons coupled antiferromagnetically to a conduction electron sea \cite{Stewart1984,Coleman2007}. The strong coupling causes collective hybridization (spin entanglement) between the two components and gives rise to a rich variety of emergent quantum phenomena such as unconventional superconductivity that defy a simple theoretical solution. Recently, it has been shown that a large amount of experimental data may be understood within the framework of a phenomenological two-fluid model \cite{Nakatsuji2004,Curro2004,Yang2008a,Yang2008b,Yang2012,Yang2013,Yang2014a,Yang2014b}. In this model, the two-component system is approximately described by two coexisting fluids: one fluid of itinerant electrons that become heavy due to the collective hybridization, and one fluid of residual unhybridized local moments whose strength is reduced accordingly. The two fluids can be viewed as the renormalized counterparts of the original two components due to the Kondo coupling, as illustrated in \fref{Fig1}. What is usually neglected in this description is the background unhybridized conduction electrons (a third fluid) that contribute little to the thermodynamic properties but may play a major role in electron transport.

The heavy electron fluid is a composite state of the hybridized local moments and conduction electrons. The key of the two-fluid description involves a transfer of the f-electron spectral weight from the local moment component to the itinerant heavy electrons with decreasing temperature. This leads to an important concept, the hybridization "order" parameter, $f(T)$, that characterizes the fraction of the f-electron spectral weight in the heavy electron component. Detailed experimental analysis shows that it has a universal temperature dependence \cite{Yang2008a},
\begin{equation}
f(T)=f_0\left(1-\frac{T}{T^*}\right)^{3/2},
\label{Eq1}
\end{equation}
where $f_0$ is the hybridization effectiveness controlling the efficiency of the collective hybridization, and $T^*$ is the coherence temperature marking the onset of the process. Both parameters may vary with pressure, magnetic field, doping or other external tuning parameters. We see that heavy electrons emerge gradually as temperature falls below $T^*$. The value of $f_0$ determines the fraction of the two components at low temperatures and therefore the properties of the ground state: for $f_0>1$, $f(T)$ approaches unity at a finite temperature $T_L$ so that all f-electrons become itinerant below $T_L$ and one may obtain a Fermi liquid state at lower temperatures; for $f_0<1$, a fraction of the local moments may persist down to very low temperatures and become magnetically ordered; $f_0=1$ thus marks a crossover or a phase transition between these two states and the system is located at a quantum critical point (QCP) at $T=0$.

The two-fluid model simplifies the complicated Kondo lattice problem to a problem of two interacting fluids. Each fluid can be given an approximate description based on experimental analysis or simple theoretical considerations \cite{Yang2008a,Yang2012}. It can be shown that the itinerant heavy electron fluid, hereafter the Kondo liquid, has an effective mass that diverges logarithmically with temperature \cite{Yang2008a},
\begin{equation}
\frac{m^*}{m_0}\sim 1+\ln\frac{T^*}{T}.
\label{Eq2}
\end{equation}
Combing this with $f(T)$ in \eref{Eq1} yields a universal density of states for the Kondo liquid,
\begin{equation}
\rho_{KL}(T)\propto f_0\left(1-\frac{T}{T^*}\right)^{3/2}\left(1+\ln\frac{T^*}{T}\right),
\label{Eq3}
\end{equation}
which is independent of the material details and indicates that the heavy electron Kondo liquid is a new quantum state of matter protected by some universal properties of the Kondo lattice. The above formulas were first derived through a combined analysis of the specific heat and the magnetic susceptibility of the La-doped CeCoIn$_5$ \cite{Yang2008a}, using the two-fluid equations presented later in this article. It is suggested that the Kondo liquid may have a constant Wilson ratio \cite{Nakatsuji2004} so its magnetic specific heat follows the same scaling. This universality has been examined in a large amount of experimental analyses \cite{Yang2012,Yang2013,Yang2014a,Yang2014b}. We will show that it is best observed in the Knight shift measurement and the Hall effect \cite{Yang2008a}. 

The local moment fluid, on the other hand, is material dependent. It can be described as a lattice of interacting spins with reduced strength of $f_l(T)=1-f(T)$. We call this a hybridized spin liquid. In the mean field approximation, it has a magnetic susceptibility \cite{Yang2012},
\begin{equation}
\chi_l(\mathbf{q})=\frac{f_l\chi_0}{1-J_{\mathbf{q}}f_l\chi_0},
\label{Eq4}
\end{equation}
in which $\chi_0$ is the susceptibility of individual local moment and $J_{\mathbf{q}}$ is the $\mathbf{q}$-dependent exchange coupling between local moments that depends on the material details.

As we will show, the two-fluid model provides a unified explanation to a number of normal state properties of heavy electron materials (see Supplementary Information in \cite{Yang2008b}). Among them, the emergence of heavy electrons is seen from the opening and rapid development of the hybridization gap in the optical conductivity \cite{Singley2002}, the growth of the quasiparticle peak in the angle-resolved photoemission spectroscopy (ARPES) \cite{Mo2012}, the Fano line shape in the point contact \cite{WPark2008} and scanning tunneling spectroscopies \cite{Aynajian2010} and the Raman spectroscopy \cite{Martinho2007}, as well as an anomalous temperature dependence in the Hall coefficient \cite{Hundley2004}. The corresponding loss of strength of the local moments is manifested in the deviation of the magnetic susceptibility from the Curie-Weiss law, the change of slope in the nuclear magnetic resonance (NMR) spin-lattice relaxation rate as a function of temperature, and the coherence peak in the magnetic resistivity due to the suppression of the Kondo scattering. We will discuss some of these in detail in the next section.

These different properties reflect the different aspects of the heavy electron physics. The observation that all of them occur at approximately the same temperature, $T^*$, provides a strong support for a common origin, namely the heavy electron emergence accompanied with the loss of the local moment spectral weight as stated in the two-fluid model \cite{Yang2008b}. It can be imagined that even a simple combination of the two distinct fluids will lead to rather complicated behaviors. This is the primary cause for the puzzling temperature evolution of many lattice properties that defy a simple theoretical understanding for four decades. 

The two-fluid model provides an indispensible way to disentangle these two coexisting components. Its implication on the microscopic theory of heavy electron physics will be discussed in another article in this issue \cite{Lonzarich2016}. Here we focus on its implementation in experiment. We will show that it can cover a large number of experimental data. We start first with the normal state properties (section 2) and then discuss how the model may be extended to understand the various low temperature ordered states (section 3).

\section{Normal state properties}
Heavy electron materials exhibit a number of anomalous properties in the normal states that demand a unified explanation. The two-fluid model relates most of these anomalies to the emergence of the heavy electron Kondo liquid and the corresponding loss of strength of the local moments below $T^*$. In this section, we will discuss some of these anomalous properties and show that the simple two-fluid model could give a quantitative explanation to various experiments. We discuss first the magnetic, thermal, transport and spectroscopic properties of heavy electron materials. We will show that these reveal different aspects of the two fluids. We then discuss the microscopic origin of the temperature scale $T^*$ and show that it leads to new insights and a different perspective on the true nature of heavy electron physics.

\subsection{Magnetic properties}
We show in this section that the NMR Knight shift and the spin-lattice relaxation rate provide the most evident experimental support for the two-fluid model, while the magnetic susceptibility provides a simple illustration on the role of the hybridization parameter, $f(T)$.

\subsubsection{The NMR Knight shift}

\begin{figure}[t]
\centerline{{\includegraphics[width=0.5\textwidth]{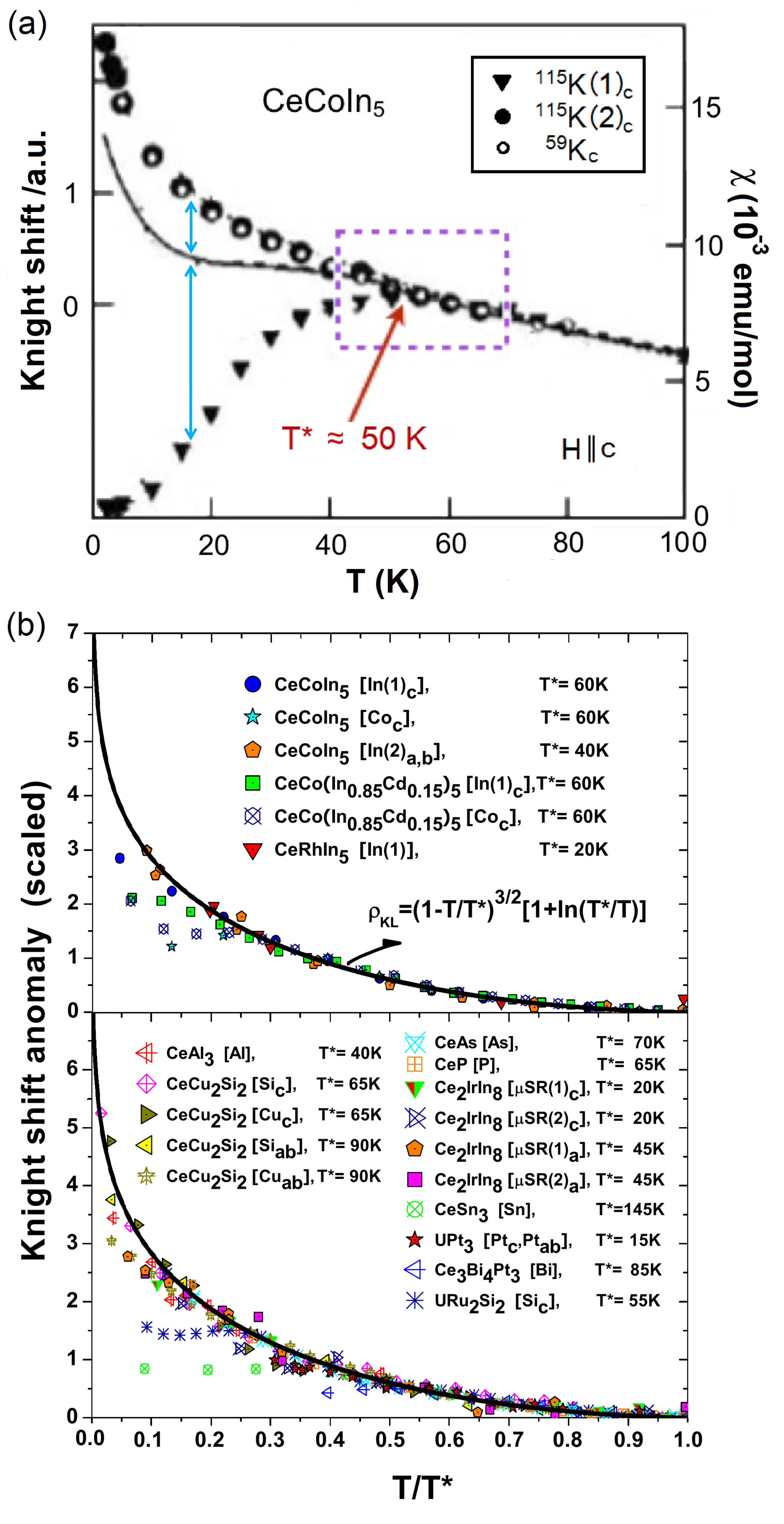}}}
\caption{
{(Color online) (a) Comparison of the $c$-axis Knight shift and the magnetic susceptibility (solid line) in CeCoIn$_5$. The two follow each other above $T^*\sim50\,$K but start to deviate below $T^*$, which defines the Knight shift anomaly. Figure adapted with permission from \cite{Curro2001}. (b) The Knight shift anomaly as a function of $T/T^*$ in a dozen of heavy electron materials, showing universal temperature scaling independent of material details. Figure adapted from \cite{Yang2008a}. }
\label{Fig2}}
\end{figure}

The NMR Knight shift originates from the hyperfine coupling between the probe nuclei and the surrounding electrons polarized by an external magnetic field. For a simple metal, it is typically proportional to the magnetic susceptibility of the conduction electrons. For heavy electron materials, the proportionality also holds above the coherence temperature, $T^*$, where localized f-moments dominate the magnetic properties and conduction electrons only contribute a small constant background. Below $T^*$, however, this simple relation fails and the Knight shift and the susceptibility exhibit an anomalous deviation from each other, as shown in \fref{Fig2}(a) for CeCoIn$_5$ \cite{Curro2001}. This has often been attributed to crystal field effects in the literature. However, as shown in \fref{Fig2}(b), detailed analysis of a dozen of heavy electron compounds indicates that this anomalous deviation exhibits universal temperature dependence, regardless of material details, and cannot be due to crystal field effects \cite{Yang2008a}. 

In the two-fluid framework, this anomaly is taken as an evidence for the emergence of the heavy electron Kondo liquid, which has a different hyperfine coupling compared to that of the local moments. The Knight shift and the magnetic susceptibility are then given by \cite{Yang2008a}
\begin{eqnarray}
\chi&=&f(T)\chi_{KL}+\left[1-f(T)\right]\chi_{SL},\\
K&=&K_0+Af(T)\chi_{KL}+B\left[1-f(T)\right]\chi_{SL},
\label{Eq5&6}
\end{eqnarray}
where $\chi_{KL}$ and $\chi_{SL}$ are the intrinsic magnetic susceptibilities of the Kondo liquid and the hybridized spin liquid, respectively;  $A$ and $B$ are their hyperfine couplings. For $T>T^*$, only local moments exist and one recovers the linear relation between the two quantities: $K=K_0+B\chi$; whereas for $T<T^*$, the difference in the hyperfine couplings $A$ and $B$ leads to an anomalous deviation,
\begin{equation}
K_{anom}=K-K_0-B\chi=(A-B)f(T)\chi_{KL}.
\label{Eq7}
\end{equation}
The Knight shift anomaly therefore probes the intrinsic susceptibility, or the density of states, of the emergent heavy electron Kondo liquid. As shown in \fref{Fig2}(b), the subtracted results are in good agreement with the predicted scaling of the Kondo liquid. The fact that all these data collapse on a single universal curve in a broad range of temperature implies that the Kondo liquid emergence has a common mechanism that is independent of material details. This universal scaling has been examined in more recent experiments \cite{Ohishi2009} and newly discovered compounds \cite{Warren2011}. It has not been expected in all other theories and therefore represents the most unique feature of the two-fluid model.

\subsubsection{The NMR spin-lattice relaxation rate}

According to the Moriya formula \cite{Moriya1956}, the spin-lattice relaxation rate is related to the imaginary part of the dynamic susceptibility,
\begin{equation}
\frac{1}{T_1}=\gamma^2T\lim_{\omega\rightarrow 0}\sum_{\mathbf{q}}F(\mathbf{q})^2\frac{\rm{Im}\chi(\mathbf{q},\omega)}{\omega},
\label{Eq8}
\end{equation}
where $F(\mathbf{q})$ is the form factor and $\gamma$ is the gyromagnetic ratio. It is immediately seen that the spin-lattice relaxation rate must have a similar two-fluid formalism \cite{Yang2009a},
\begin{equation}
\frac{1}{T_1}=\frac{1-f(T)}{T_{1SL}}+\frac{f(T)}{T_{1KL}},
\label{Eq9}
\end{equation}
where $T_{1SL}$ and $T_{1KL}$ are the intrinsic contributions of the two fluids. 

\begin{figure}[t]
\centerline{{\includegraphics[width=0.5\textwidth]{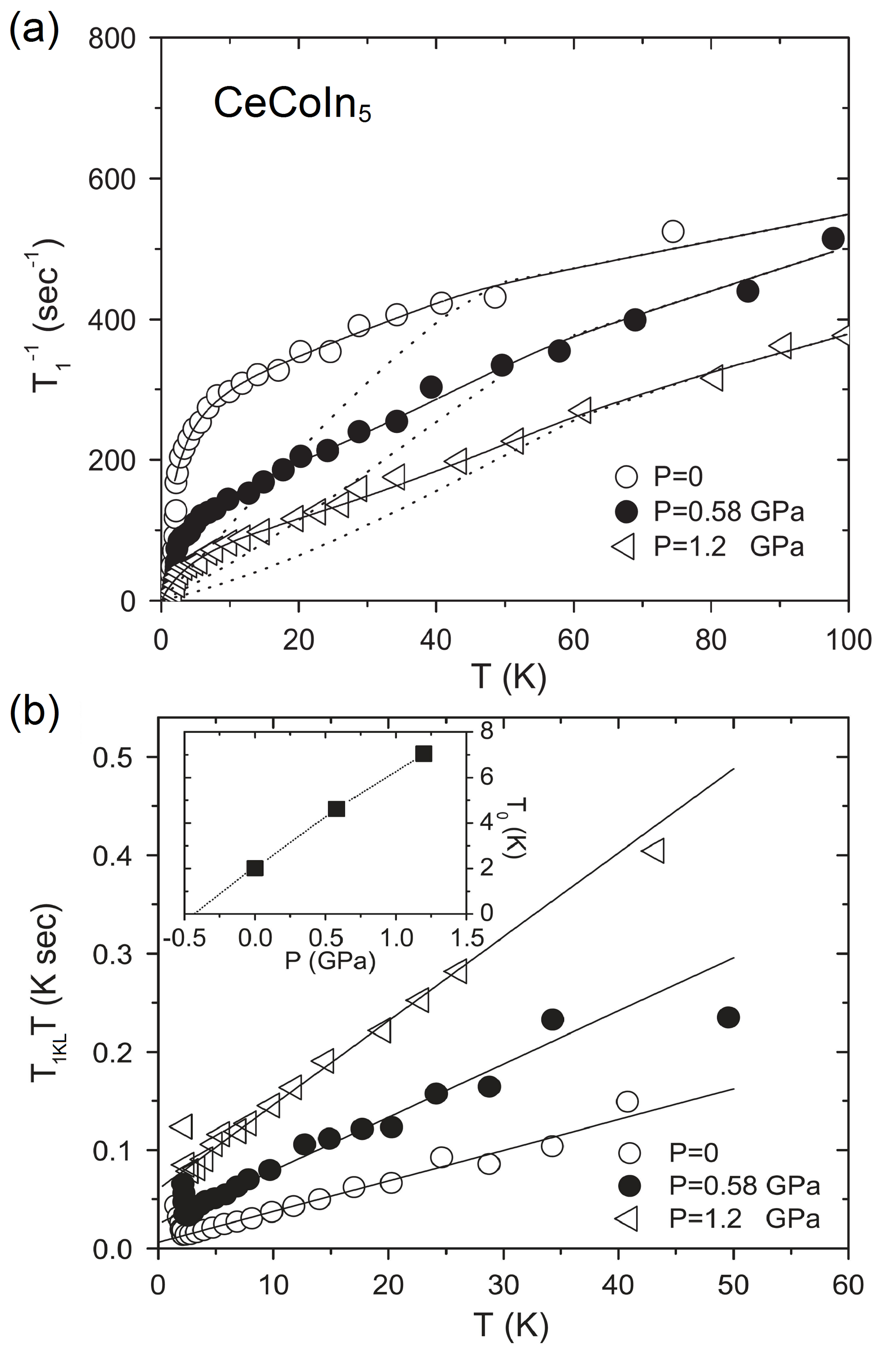}}}
\caption{
{(Color online) (a) Theoretical fit (solid lines) to the total spin-lattice relaxation rate in CeCoIn$_5$. The dashed lines are the local moment contributions. (b) Temperature dependence of the subtracted Kondo liquid $T_{1KL}$, showing $T_{1KL}T\propto(T+T_0)$. The inset plots $T_0$ as a function of pressure, extrapolating to a quantum critical point at slightly negative pressure. Figure adapted from \cite{Yang2009a}.}
\label{Fig3}}
\end{figure}

Information about $T_{1SL}$ and $T_{1KL}$ may be obtained from experimental analysis. Above $T^*$, we have $T_1=T_{1SL}$, which measures the magnetic fluctuations of the unhybridized moments. For many materials, $1/T_{1SL}$ is either constant or linear in temperature above $T^*$, as may be derived for weakly interacting local moments. If this temperature dependence of $1/T_{1SL}$ persists below $T^*$, we could then use the above two-fluid formula to subtract the Kondo liquid contribution $T_{1KL}$ and study its behavior. Detailed analysis of the spin-lattice relaxation rate has been carried out for CeCoIn$_5$ under pressure \cite{Yang2009a}. The results are plotted in \fref{Fig3}(b). Interestingly, the subtracted $T_{1KL}$ has a simple temperature dependence,
\begin{equation}
T_{1KL}T\sim (T+T_0).
\label{Eq10}
\end{equation}
Similar behavior has been observed in cuprates where it signals the presence of quantum critical fluctuations of a nearly two-dimensional (2D) spin liquid \cite{Barzykin2009}. $T_0$ measures the distance from the magnetic QCP and its pressure dependence in CeCoIn$_5$ is shown in the inset of \fref{Fig3}(b), indicating a magnetic QCP located at slightly negative pressure, as expected for CeCoIn$_5$ under high magnetic field \cite{Sidorov2002}. \Fref{Fig3}(a) compares the two-fluid fit (solid lines) to the experimental data using \eref{Eq9} and \eref{Eq10}. The dashed lines are the local moment contributions, showing large deviations from the experimental data below $T^*$.

\begin{figure}[t]
\centerline{{\includegraphics[width=0.5\textwidth]{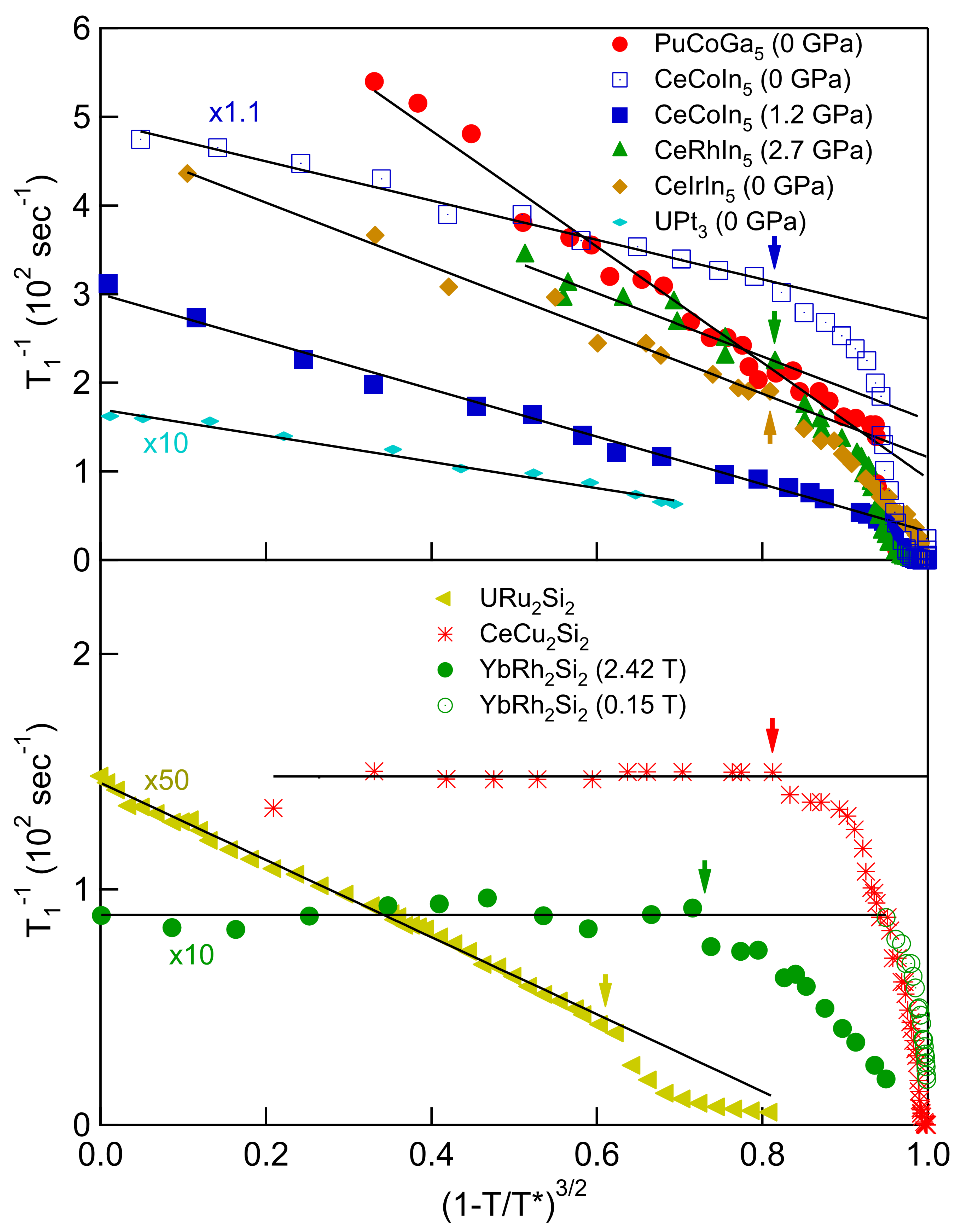}}}
\caption{
{(Color online) The inverse spin-lattice relaxation rate as a function of $(1-T/T^*)^{3/2}$, showing a universal scaling of $1/T_1$ near the magnetic QCP before it is intervened by some other low temperature physics. Figure adapted from \cite{Yang2015}.}
\label{Fig4}}
\end{figure}

The above results suggest that $T_{1KL}$ is roughly independent of temperature near the magnetic QCP. The two-fluid formula may then be rewritten as \cite{Yang2015}
\begin{equation}
\frac{1}{T_1}=\frac{1}{T_{1SL}}+\left(\frac{1}{T_{1KL}}-\frac{1}{T_{1SL}}\right)f(T).
\label{Eq11}
\end{equation}
If $1/T_{1SL}$ is also constant or only weakly temperature dependent, we may expect that $1/T_1$ exhibits a universal scaling with respect to $f(T)\propto (1-T/T^*)^{3/2}$. We have hence studied the NMR data in a number of heavy electron materials near their magnetic QCP. \Fref{Fig4} plots $1/T_1$ versus $(1-T/T^*)^{3/2}$ for some of them where $T_1$ has been measured. The nice scaling confirms the two-fluid expectation \cite{Yang2015}. \Fref{Fig4}(b) also presents the data for CeCu$_2$Si$_2$ and YbRh$_2$Si$_2$ whose $1/T_1$ are anomalously independent of temperature below $T^*$ for some unknown reason. Away from the QCP, we find relatively larger deviations from the simple scaling. This once again confirms the validity of the two-fluid model and provides an independent experimental verification for the universal scaling of the hybridization "order" parameter.

\subsubsection{The magnetic susceptibility}

\begin{figure}[t]
\centerline{{\includegraphics[width=0.8\textwidth]{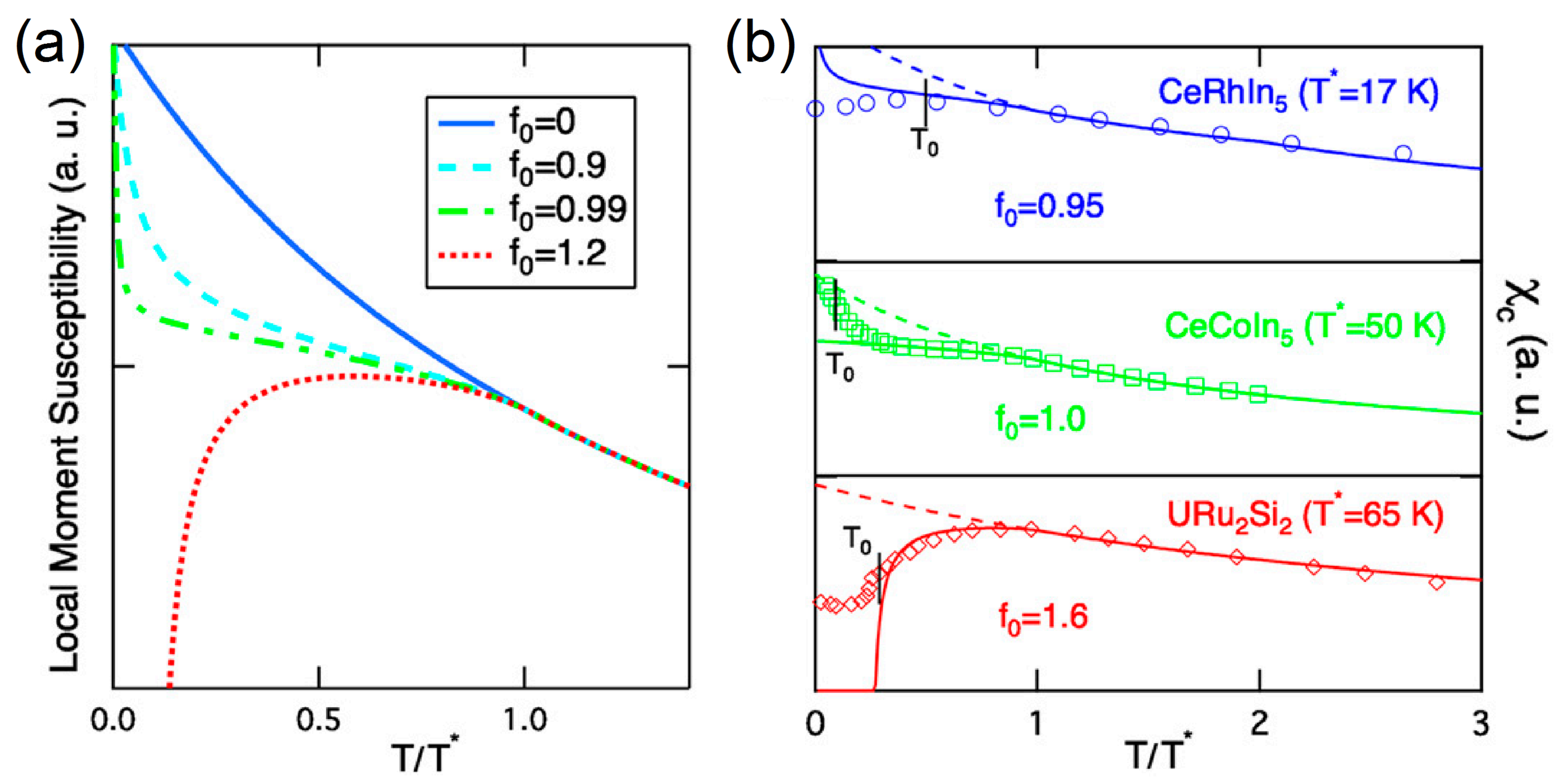}}}
\caption{
{(Color online) (a) Predicted local moment susceptibility for different values of $f_0$; (b) Fit to the experimental data with chosen values of $f_0$ for CeRhIn$_5$, CeCoIn$_5$ and URu$_2$Si$_2$. $T_0$ is the cutoff temperature below which other effects set in. Figure adapted from \cite{Yang2012}.}
\label{Fig5}}
\end{figure}

In most heavy electron materials, the magnetic susceptibility exhibits the Curie-Weiss behavior above $T^*$, which is a manifestation of fluctuating unhybridized local moments at high temperatures. The deviation from the Curie-Weiss law below $T^*$ was often attributed to crystal field effects in the literature but is understood in the two-fluid model to result from the loss of strength of the local moments due to the collective hybridization. The mean-field approximation in \eref{Eq4} allows us to study qualitatively the local moment susceptibility and its variation with the hybridization parameter, $f_0$ \cite{Yang2012}. Comparison with experimental data is plotted in \fref{Fig5} with different chosen values of $f_0$ for the local moment antiferromagnet CeRhIn$_5$, the quantum critical superconductor CeCoIn$_5$ and the more itinerant 5f-compound URu$_2$Si$_2$. For $f_0<1$, the local moment susceptibility continues to grow with decreasing temperature and deviates only slightly from the Curie-Weiss law below $T^*$ until a peak shows up as a precursor of the magnetic order at much lower temperature ($T_N$=3.8 K for CeRhIn$_5$); whereas for $f_0>1$, the susceptibility is more rapidly suppressed with a broad peak slightly below $T^*$ due to the rapid delocalization of the f-moments as in URu$_2$Si$_2$; in between for $f_0\approx1$, one sees a plateau in the susceptibility, as observed in the quantum critical superconductor CeCoIn$_5$. Hence the behavior of the magnetic susceptibility around $T^*$ provides a qualitative measure of $f_0$ which in turn determines the ordered state at low temperatures.

\subsection{Transport and electronic properties}

In this section, we discuss the transport and electronic properties including the Hall effect, the Fano interference effect in the scanning tunneling and point contact spectroscopies, and the quasiparticle peak in ARPES. These experiments reveal the very special composite nature of the emergent heavy electrons.

\subsubsection{The Hall anomaly}
The Hall coefficient in heavy electron materials is typically dominated by the skew scattering of conduction electrons off independent f-moments, 
\begin{equation}
R_H=R_0+r_l\rho_m\chi,
\label{Eq12}
\end{equation}
where $R_0$ is the ordinary Hall coefficient, $r_l$ is a constant, $\rho_m$ is the magnetic resistivity, $\chi$ is the magnetic susceptibility, and $R_s=r_l\rho_m\chi$ is the extraordinary or anomalous Hall contribution first proposed by Fert and Levy in 1987 \cite{Fert1987}. The above formula has been verified in many heavy electron materials such as CeAl$_3$ and CeCu$_2$Si$_2$ in the high temperature regime but fails when coherence sets in below $T^*$ \cite{Hadzic1986}. In the caged compound Ce$_3$Rh$_4$Sn$_{13}$, in which no lattice coherence and long-range orders is observed, the theory is found to be valid down to the lowest measured temperature \cite{Kohler2007}. However, so far no theory other than the two-fluid model allows for a quantitative analysis of the experimental data in the coherent regime.

\begin{figure}[t]
\centerline{{\includegraphics[width=0.5\textwidth]{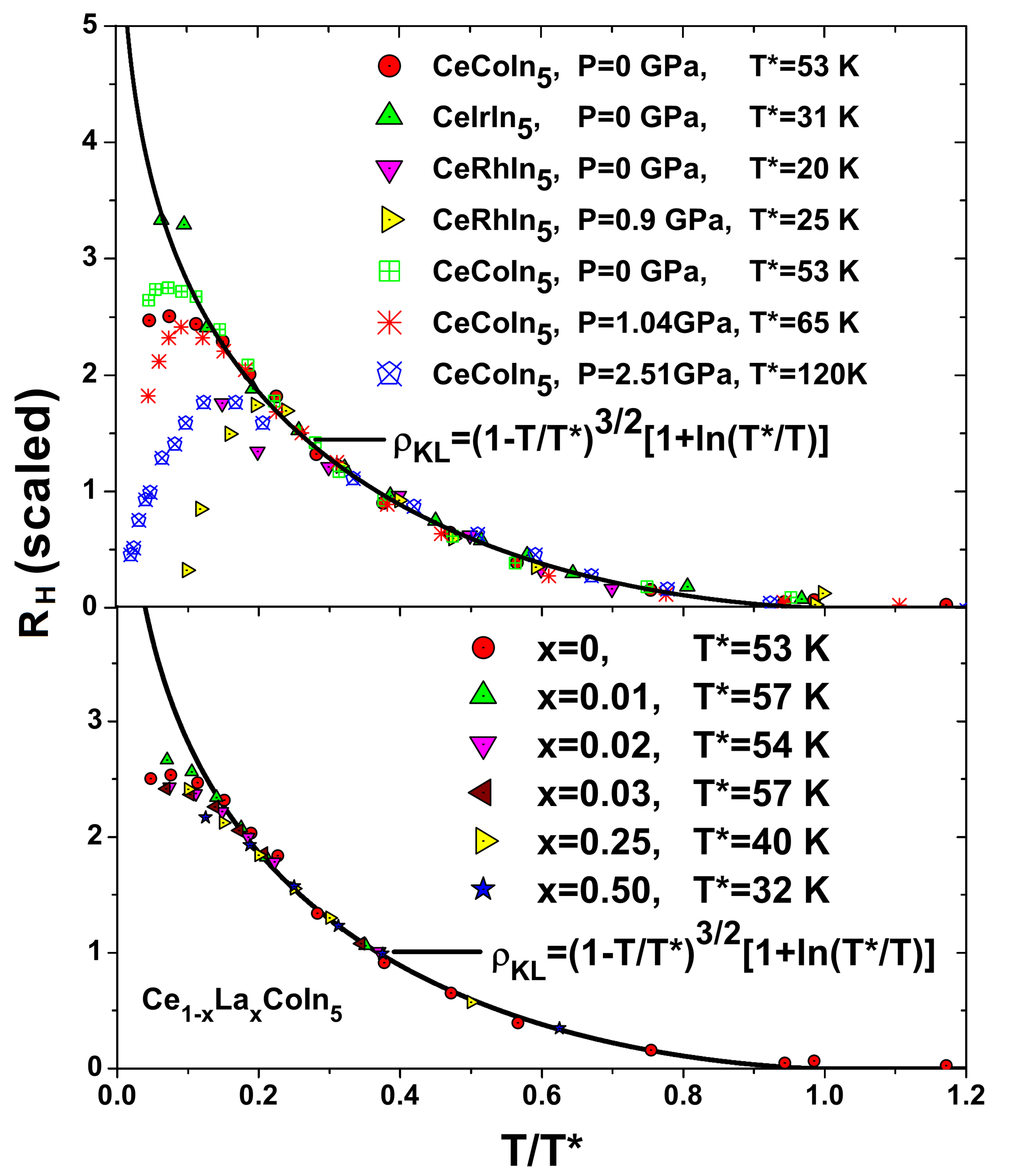}}}
\caption{
{(Color online) The Hall coefficient (scaled) as a function of temperature in Ce$M$In$_5$ under pressure and with doping, showing the universal scaling predicted by the two-fluid model. Figure adapted from \cite{Yang2008a}.}
\label{Fig6}}
\end{figure}

Important progress was first made following the observation of the puzzling behavior in the temperature dependence of the Hall coefficient in Ce$M$In$_5$ \cite{Hundley2004,Nakajima2007}. Unlike most other compounds, their Hall coefficients are almost independent of temperature above $T^*$, implying that Fert and Levy's incoherent skew scattering contribution is suppressed, namely $r_l\approx0$. However, a strong temperature dependence is developed below $T^*$, accompanying with the onset of coherence and following exactly the predicted universal temperature scaling of the Kondo liquid, as plotted in \fref{Fig6} \cite{Yang2008a} and later examined in Ce$_2$PdIn$_8$ \cite{Gnida2012} and CeIn$_3$ \cite{Araki2015}. One may therefore conclude that the heavy electrons contribute very differently to the Hall coefficient. This leads to the proposal of an empirical two-fluid formula for the Hall coefficient \cite{Yang2013},
\begin{equation}
R_H=R_0+r_l\rho_m\chi_l+r_h\chi_h,
\label{Eq13}
\end{equation}
where $r_h$ is a constant and $r_h\chi_h$ is the contribution of the Kondo liquid. $\chi_l=\left[1-f(T)\right]\chi_{SL}$ and $\chi_h=f(T)\chi_{KL}$ are the magnetic susceptibility of the two fluids, respectively. The above formula can be approximately derived if we consider the heavy electrons and the unhybridized light conduction electrons as two types of charge carriers \cite{Yang2013}. The unhybridized light conduction electrons are normally neglected in the two-fluid analysis due to their relatively small contributions to the magnetic susceptibility and the specific heat. However, their incoherent magnetic scattering off the residual local moments yields major contributions above or near $T^*$ to the transport properties including the magnetic resistivity and the Hall coefficient. For the Hall coefficient, they give rise to the skew scattering contribution proposed by Fert and Levy. On the other hand, as discussed in \cite{Kontani1994}, the heavy electrons have very different properties due to their coherent nature, whose contributions to the conductivity and the Hall coefficient grow gradually with lowering temperature and become dominant in the fully coherent regime. The fact that the two dominate in somewhat different regimes lead to the peculiar two-fluid form of \eref{Eq13}.

\begin{figure}[t]
\centerline{{\includegraphics[width=0.5\textwidth]{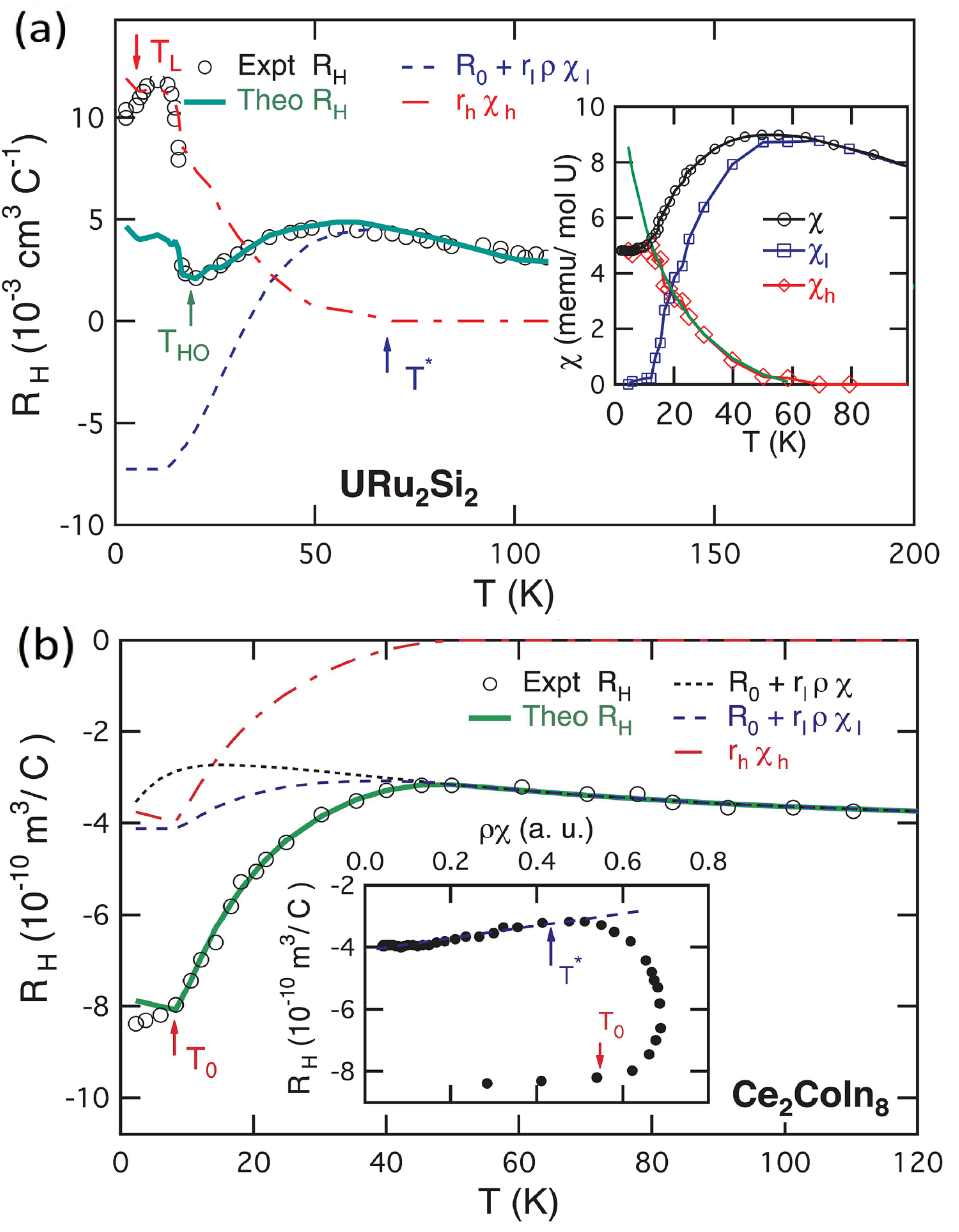}}}
\caption{
{(Color online) Two-fluid analysis of the Hall coefficient in (a) URu$_2$Si$_2$ \cite{Schoenes1987} and (b) Ce$_2$CoIn$_8$ \cite{Chen2003}. The solid lines are the overall fit and the dash-dotted line indicates a significant contribution from the emergent heavy electrons. The two insets show the magnetic susceptibility of the two fluids in URu$_2$Si$_2$ and the deviation of $R_H$ from the Fert-Levy formula in Ce$_2$CoIn$_8$ below $T^*$, respectively. Figure adapted from \cite{Yang2013}.}
\label{Fig7}}
\end{figure}

The above formula for the Hall coefficient provides a simple interpolation between the two limits. For $T>T^*$, it reduces to the usual Fert-Levy formula, while in the limit $r_l=0$, it yields the Kondo liquid scaling, $R_H=R_0+r_h\chi_h$. Its validity has been examined in more general cases. Using $\chi_l$ and $\chi_h$ obtained from combined analysis of the susceptibility and the Knight shift data or simply from the scaling formula of the Kondo liquid, we have applied \eref{Eq13} to URu$_2$Si$_2$ \cite{Schoenes1987} and Ce$_2$CoIn$_8$ \cite{Chen2003}. The constants $R_0$ and $r_l$ can both be determined from high temperature fit above $T^*$, so that only one free parameter $r_h$ is left to fit the whole temperature evolution below $T^*$. Detailed analysis can be found in \cite{Yang2013} and the results are shown in \fref{Fig7}. The excellent agreement in a wide temperature range for both compounds confirms the proposed empirical formula. In both cases, we see that the Kondo liquid contributes a considerable part of the total Hall coefficient.

The two-fluid formula of $R_H$ is a result of the changing character of the f-electrons from localized moments to itinerant heavy electrons. It allows for a consistent interpretation as well as a better data analysis of the Hall experiment over a broad temperature range. Further investigations are crucial in order to achieve a thorough understanding of its validity.

\subsubsection{The Fano interference}

\begin{figure}[t]
\centerline{{\includegraphics[width=0.5\textwidth]{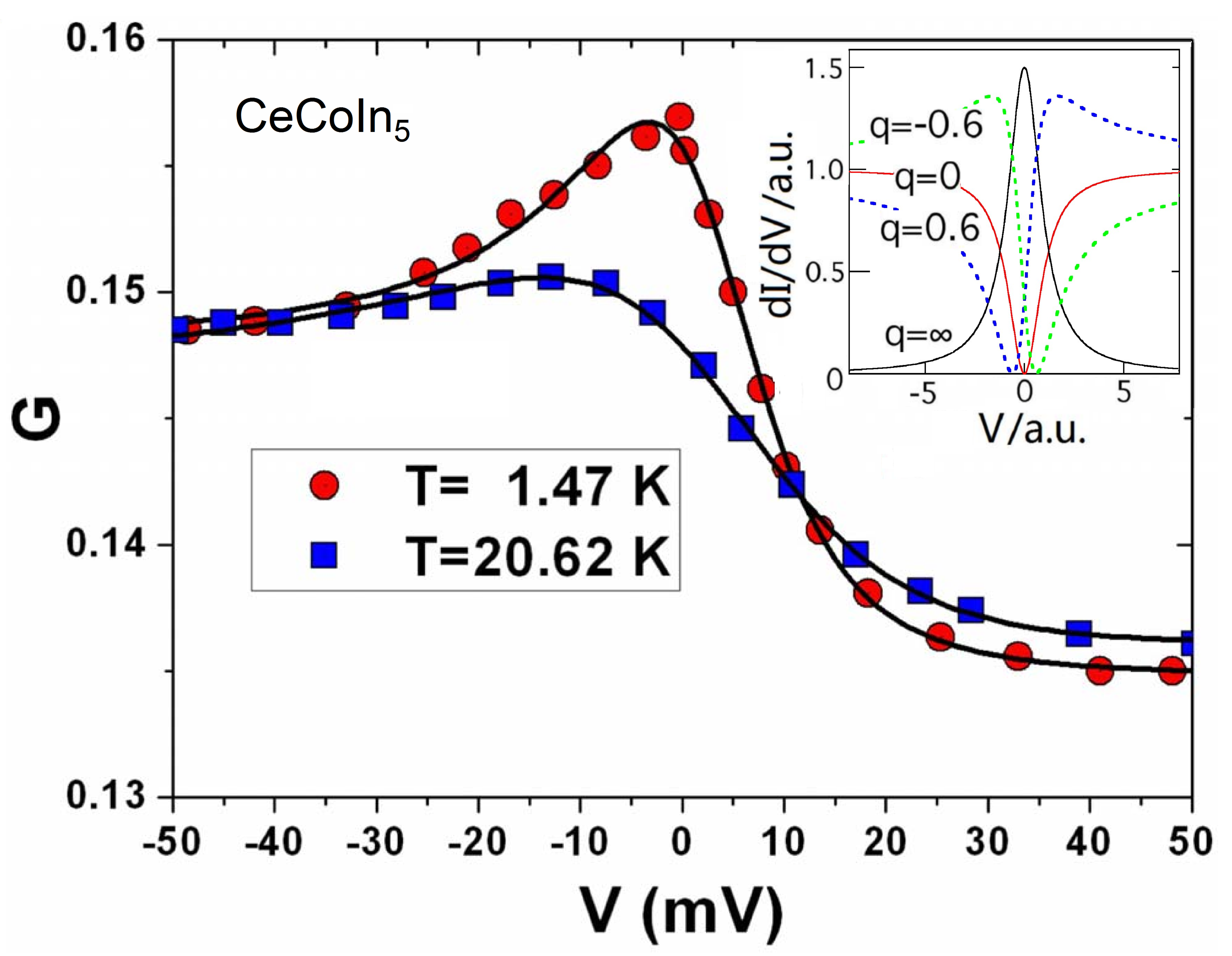}}}
\caption{
{(Color online) The Fano fit to the point-contact spectra of CeCoIn$_5$. The inset shows the typical Fano line shape for different values of $q$. Figure adapted from \cite{Yang2009b}.}
\label{Fig8}}
\end{figure}

Important information on the nature of the emergent heavy electrons can be obtained from the point contact spectroscopy (PCS) and the scanning tunneling spectroscopy (STS), which exhibit asymmetric differential conductance at large positive and negative bias voltages \cite{WPark2008}. This was first explained theoretically by the author based on the interference effect of the tip electrons injecting simultaneously into the conduction and f-electron channels \cite{Yang2009b}. Since then, a number of different approaches depending on the approximation for the Kondo lattice have been applied to the problem and yielded similar results \cite{Maltseva2009,Wolfle2009,Fogelstrom2010}. In the mean-field approximation, we can derive a simple formula for the conductance \cite{Yang2009b},
\begin{equation}
G(V)=g_0+\int_{-\infty}^\infty dE\, g_I(E)T(E)\frac{df_{FD}(E-eV)}{d(eV)}\approx g_0+g_IT(eV),
\label{Eq14}
\end{equation}
with
\begin{equation}
T(E)=\frac{\left|q-\tilde{E}\right|^2}{1+\tilde{E}^2},
\label{Eq15}
\end{equation}
in which $V$ is the bias voltage, $g_0$ and $g_I$ are both constants, $f_{FD}(E)$ is the Fermi distribution function, and $T(E)$ has the Fano line shape originating from the hybridization between the broad conduction electron band and the narrow f-electron band. We have defined $\tilde{E}=(E-\epsilon_0)/\tilde{V}$, where $\epsilon_0$ is the renormalized f-electron energy and $\tilde{V}$ is the effective hybridization between the two bands. The Fano parameter, $q$, is given by the ratio of the tunneling elements between the f- and conduction channels. It determines the overall line shape of the spectra, as illustrated in the inset of \fref{Fig8}. This prediction of the Fano interference has now been verified in a number of compounds such as CeCoIn$_5$ \cite{WPark2008}, URu$_2$Si$_2$ \cite{Aynajian2010,Park2012}, and SmB$_6$ \cite{Zhang2013,Rossler2014}. As an example, \fref{Fig8} shows the Fano fit to the PCS data in CeCoIn$_5$. The results provide a clear demonstration of the hybridization physics in heavy electron materials and reveal the composite nature of the emergent heavy electrons.

An important issue that has not been widely discussed in the literature is the difference in the observed conductance spectra of PCS and STS. The distinction reflects the fundamental difference between the usual Fano systems and the Kondo lattice and may be seen in the above formula through the energy-dependent prefactor,
\begin{equation}
g_I(E)\propto\rho_t\sum_{ikm}\left|M_{ckm}\right|^2\delta\left(E-\epsilon_{ik}\right),
\label{Eq16}
\end{equation}
which is a convolution of the electronic band structure and the tunneling matrix, $M_{ckm}$, between the tip and the conduction channel. $\rho_t$ is the density of states of the tip and $\epsilon_{ik}$ is the dispersion of the $i$-th hybridization band of the Kondo lattice. For STS, the tip is local in space so that $M_{ckm}$ is $k$-independent so that the prefactor $g_I(E)$ is proportional to the total density of state of the heavy electrons; while for PCS, $M_{ckm}$ is $k$-dependent and $g_I(E)$ involves a weighted average in the momentum space. As a result, STS shows clear signature of the hybridization gap, whereas in PCS the hybridization gap is often smeared out and one sees only the Fano line shape.

\subsubsection{ARPES}

\begin{figure}[t]
\centerline{{\includegraphics[width=0.5\textwidth]{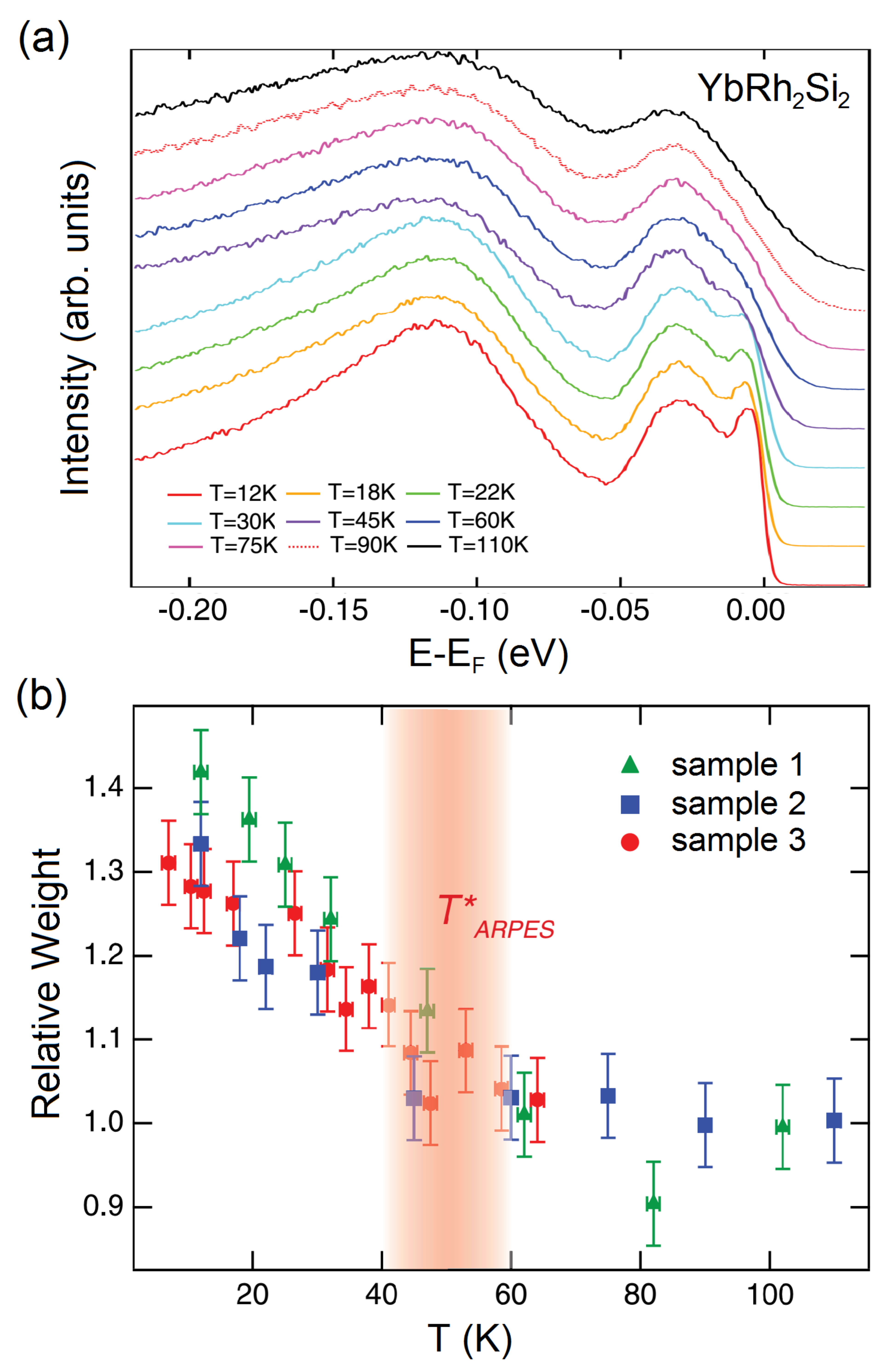}}}
\caption{
{(Color online) (a) ARPES data of YbRh$_2$Si$_2$ as a function of temperature; (b) Temperature dependence of the quasiparticle spectral weight. Figure adapted from \cite{Mo2012}.}
\label{Fig9}}
\end{figure}

ARPES provides direct evidence for the emergence of heavy electrons below $T^*$. \Fref{Fig9} reproduces the experimental results for YbRh$_2$Si$_2$ \cite{Mo2012}. We see the gradual growth of a quasiparticle peak near the Fermi energy. The increase of the f-electron spectral weight with decreasing temperature is consistent with the prediction of the two-fluid model. The onset temperature, $\sim50\,$K, also agrees with the deduced value of $T^*$ from various magnetic, thermal and transport measurements \cite{Yang2008b}. This provides a microscopic justification for the two-fluid scenario, namely the emergence of heavy electrons below $T^*$. We note that the observation of such a temperature variation of the f-electron spectral weight represents a tremendous experimental progress. It has not been possible due to the energy resolution of the ARPES experiment. The lack of its observation previously was in contradiction with theoretical expectations and has led to considerable confusions.

\subsection{Thermal properties}

The emergence of heavy electrons is accompanied with the suppression of the magnetic entropy, indicating the importance of spin entanglement. The two-fluid model allows us to make quantitative predictions on the temperature dependence of the magnetic entropy \cite{Yang2012},
\begin{equation}
S(T)=\left[1-f(T)\right]S_{SL}(T)+f(T)S_{KL}(T),
\label{Eq17}
\end{equation}
where $S_{SL}$ is the intrinsic entropy of the local moments and may be approximated as $R\ln2$ for weakly interacting moments ($R$ is the gas constant), and $S_{KL}$ is the intrinsic entropy of the heavy electrons. The specific heat coefficient is then
\begin{equation}
\frac{C}{T}=\frac{dS}{dT}=\left[1-f(T)\right]\frac{C_{SL}}{T}+f(T)\frac{C_{KL}}{T}+\frac{df(T)}{dT}\left(S_{KL}-S_{SL}\right),
\label{Eq18}
\end{equation}
in which the third term involves the change in the f-electron spectral weight of the two fluids and was not included in previous analysis of La-doped CeCoIn$_5$ \cite{Nakatsuji2004}. If we assume that the Kondo liquid has a constant Wilson ratio, its specific heat coefficient should exhibit the same scaling,
\begin{equation}
\frac{C_{KL}}{T}\propto \left(1+\ln\frac{T^*}{T}\right),
\label{Eq19}
\end{equation}
which can be integrated to give the entropy \cite{Yang2012},
\begin{equation}
S_{KL}(T)=R\ln2\frac{T}{2T^*}\left(2+\ln\frac{T^*}{T}\right),
\label{Eq20}
\end{equation}
where the prefactor is determined such that $S_{KL}(T^*)=R\ln2$ for materials with a ground state doublet. Comparisons with experimental data will be discussed in section 3.1 for a number of heavy electron materials with nonmagnetic ground state. For the antiferromagnet CeRhIn$_5$ with $T_N=3.8\,$K and $T^*\approx 17\,$K, the fraction of unhybridized local moments can be estimated using $f_l(T_N)\approx 0.32$, which is consistent with the experimental observation of 30\% entropy release at $T_N$. This agreement confirms that not all f-electrons get ordered at $T_N$ as expected in the two-fluid model and indicates the correlated nature of the normal state heavy electrons.

\subsection{Origin of $T^*$}

\begin{figure}[t]
\centerline{{\includegraphics[width=0.5\textwidth]{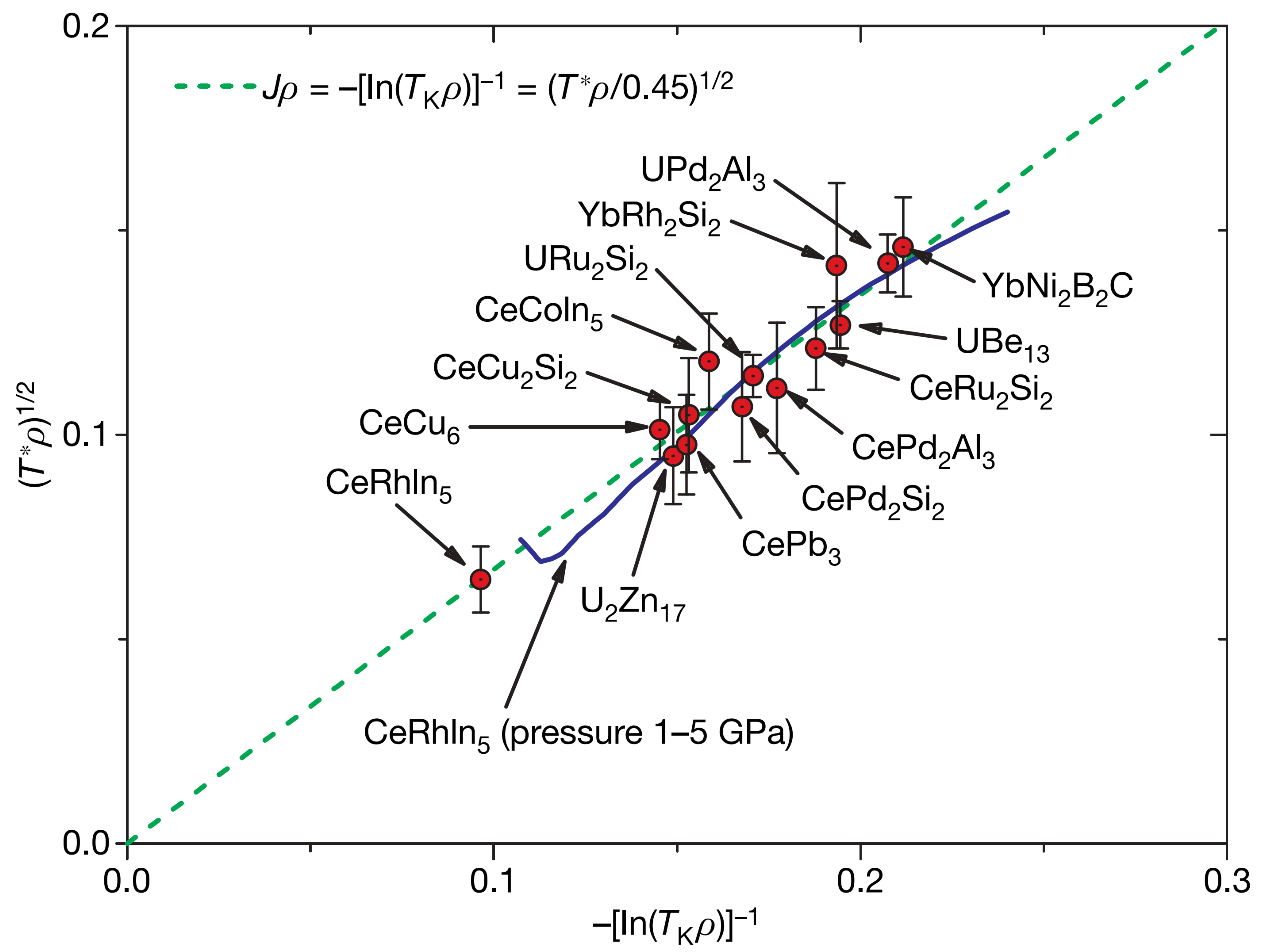}}}
\caption{
{(Color online) Comparison between $T^*$ and the single ion Kondo temperature, $T_K$, that confirms the RKKY origin of $T^*$. Figure adapted from \cite{Yang2008b}.}
\label{Fig10}}
\end{figure}

The success of the two-fluid model demands a microscopic understanding of its underlying mechanism. The first question concerns the origin of the characteristic temperature, $T^*$, that governs the onset of the two-fluid behavior. Since the Kondo coupling is the basic interaction in the system, it is natural to ask how $T^*$ may be related to the Kondo coupling, $J$. For this, we consider the diluted limit, which usually exhibits well-defined single ion Kondo behavior so that $J$ can be estimated from the measured Kondo temperature using \cite{Hewson1993},
\begin{equation}
T_K=\rho^{-1}\rme^{-1/J\rho},
\label{Eq21}
\end{equation}
and
\begin{equation}
\rho=\frac{3\gamma}{\pi^2},
\label{Eq22}
\end{equation}
where $\rho$ is the density of states of the background conduction electrons whose value may be estimated from the specific heat coefficient $\gamma$ of the nonmagnetic host (e.g., LaCoIn$_5$ for CeCoIn$_5$). The Boltzmann constant $k_B$ is set to unity. Following the analysis in \cite{Yang2008b}, we can estimate the magnitude of $J$ that governs the Kondo lattice after proper volume corrections. \Fref{Fig10} compares the values of $T^*$ and $T_K$ in a number of heavy electron compounds, including CeRhIn$_5$ under pressure (1-5 GPa). We find that
\begin{equation}
T^*=cJ^2\rho,
\label{Eq23}
\end{equation}
where $c\approx0.45$ is a constant. This indicates that $T^*$ is given by the Ruderman-Kittel-Kasuya-Yosida (RKKY) interaction between neighboring f-moments, as previously observed in Ce$_{1-x}$La$_x$CoIn$_5$ \cite{Nakatsuji2002}. In fact, many heavy electron materials that exhibit quantum critical behavior appear to cluster between $J\rho=0.15$ and 0.20, where $T^*$ is much greater than the single ion Kondo temperature. This suggests that heavy electron physics is a genuine lattice effect and cannot be viewed as a simple lattice extension of the Kondo physics. This observation is in radical contradiction with the conventional wisdom in which $T^*$ is attributed to the Kondo temperature, $T_K$. We also note that the prefactor $c$ seems to be universal for a broad range of materials that have cubic, tetragonal or hexagonal crystal structures and a magnetically ordered, superconducting, or paramagnetic ground state. This universality and the dominance of the RKKY interaction point to a completely new perspective on heavy electron physics.

\section{Low temperature states}

\begin{figure}[t]
\centerline{{\includegraphics[width=0.5\textwidth]{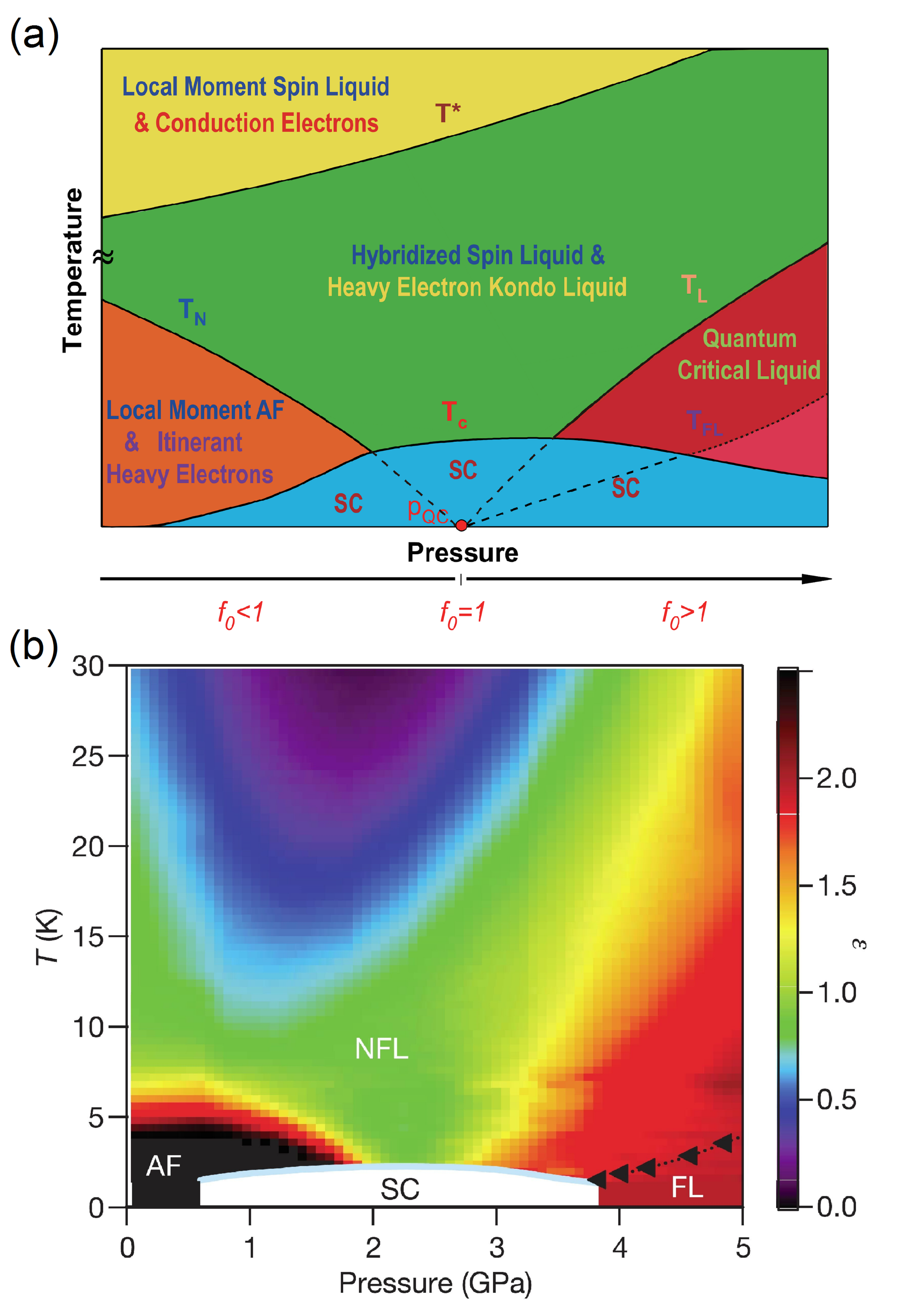}}}
\caption{
{(Color online) Comparison between (a) the predicted phase diagram of the two-fluid model and (b) the experimental phase diagram of CeRhIn$_5$. Figure adapted from \cite{Yang2012,Yang2015,TPark2008}.}
\label{Fig11}}
\end{figure}

We have shown that the two-fluid model is quite successful in explaining the normal state properties of heavy electron materials. To extend it to the low temperature ordered states, we need to consider the instabilities of both the itinerant heavy electrons and the residual local moments. This immediately leads to several important observations as illustrated in the $T-f_0$ phase diagram in \fref{Fig11}(a) \cite{Yang2012,Yang2015}:
\begin{itemize}
\item For $f_0>1$, there exists a finite temperature $T_L$, at which $f(T)$ reaches unity, that marks the complete delocalization of all f-electrons, and a Fermi liquid state may then be stabilized at a lower temperature, $T_{FL}$.
\item For $f_0<1$, a fraction of the local moments may persist down to zero temperature and give rise to a spin liquid or a magnetically ordered state.
\item For $f_0=1$, $T=0$ marks a magnetic and delocalization QCP, accompanying with a change in the Fermi surface across this point.
\end{itemize}
The experimental phase diagram of CeRhIn$_5$ is shown in \fref{Fig11}(b) for comparison \cite{TPark2008}. The overall agreement suggests that the two-fluid model is a candidate scenario for heavy electron physics at all temperatures. The overlap between the two-fluid regime in \fref{Fig11}(a) and the non-Fermi liquid regime in \fref{Fig11}(b) is a strong indication that the latter may be understood from the coexistence of the two fluids. However, detailed analysis has yet to be worked out in order to derive the unusual non-Fermi liquid scaling from the two-fluid model. Below we discuss the different low temperature regions in the phase diagram and make quantitative predictions on the ordered states.

\subsection{The Fermi liquid}

The right part ($f_0>1$) of the phase diagram in \fref{Fig11}(a) represents one of the unique features of the two-fluid model, namely the existence of a new temperature scale, the delocalization temperature $T_L$, below which all f-electrons become itinerant. $T_L$ is related to the hybridization effectiveness through $f(T_L)=1$ \cite{Yang2012}, giving
\begin{equation}
T_L=T^*\left(1-f_0^{-2/3}\right).
\label{Eq24}
\end{equation}
Below $T_L$, the coupling between the electrons and the quantum critical or Fermi surface fluctuations may lead to a region of anomalous Fermi liquid; the Fermi liquid state with well-defined Landau quasiparticles may only be realized at lower temperatures below $T_{FL}$, as shown in \fref{Fig11}(a). The delocalization line extrapolates to a delocalization QCP at $f_0=1$.

Identification of the delocalization line as a function of external parameters such as pressure or magnetic field provides a crucial test of the model. It could also yield important information on the evolution of $f_0$. Candidate signatures to be examined in future experiment may include:
\begin{itemize}
\item Fermi surface change across the delocalization line and the QCP at $T_L=0$, as observed in CeRhIn$_5$ \cite{Shishido2006} and YbRh$_2$Si$_2$ \cite{Friedemann2010};
\item Maximum in the magneto-resistivity due to density fluctuations associated with $T_L$, as observed in CeCoIn$_5$ \cite{Zaum2011};
\item Crossover behavior in the Hall coefficient as seen in YbRh$_2$Si$_2$ \cite{Paschen2004};
\item Recovery of one component behavior in the Knight shift versus the magnetic susceptibility due to the suppression of the local moment component below $T_L$.
\end{itemize}
We emphasize that detecting the Fermi surface change is an important issue in heavy electron physics. While an abrupt Fermi surface change has been observed across the delocalization QCP in several materials, it remains unclear how the Fermi surface may actually evolve with temperature. 

\begin{figure}[t]
\centerline{{\includegraphics[width=0.5\textwidth]{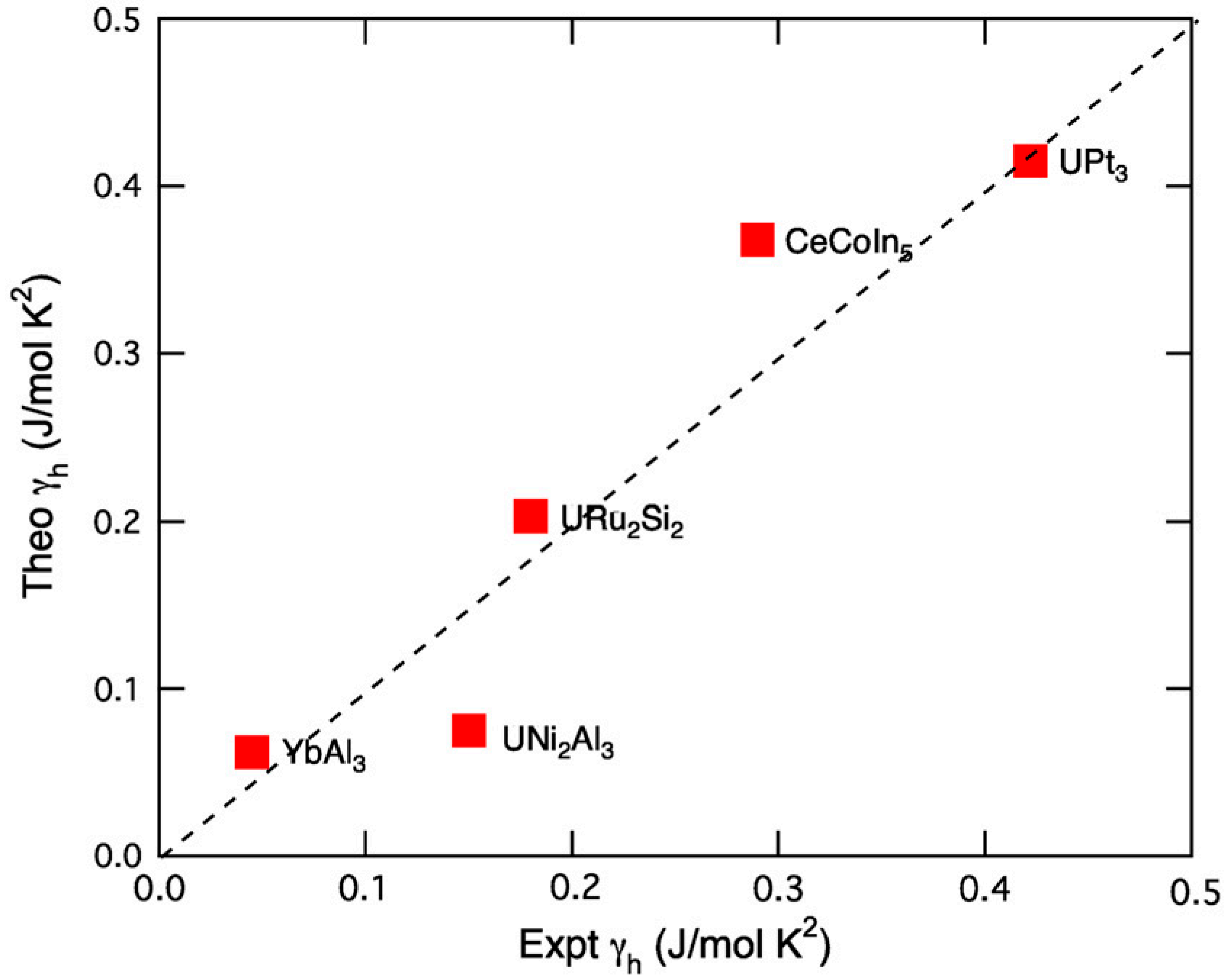}}}
\caption{
{(Color online) Comparison of experimental specific heat coefficient with the two-fluid prediction in several heavy electron compounds. Figure adapted from \cite{Yang2012}.}
\label{Fig12}}
\end{figure}

Calculations of the specific heat coefficient are greatly simplified in the Fermi liquid regime. Assuming that the specific heat coefficient is constant below $T_L$, we have from \eref{Eq20},
\begin{equation}
\gamma_{h}\approx\frac{S_{KL}(T_L)}{T_L}=\frac{R\ln2}{2T^*}\left(2+\ln\frac{T^*}{T}\right).
\label{Eq25}
\end{equation}
\Fref{Fig12} gives the predicted specific heat coefficient for a number of heavy electron materials with nonmagnetic ground states \cite{Yang2012}. The good agreement with the experimental data provides a further support for the two-fluid prediction.

\subsection{Magnetic order}

For $f_0<1$, magnetic instabilities in the residual local moments could give rise to long-range magnetic orders at low temperature. Using the simple mean-field formula \eref{Eq4} for the local moment susceptibility, we can estimate the N\'eel temperature \cite{Yang2012},
\begin{equation}
\frac{T_N(p)}{T^*(p)}=\eta_Nf_l(T_N,p),
\label{Eq26}
\end{equation}
where $\eta_N=CJ_Q/T^*$ includes the effect of frustration, while $f_l(T_N,p)$ accounts for the reduction in the local moment strength due to collective hybridization. Assuming that the scaling formula of $f(T)$ holds down to zero temperature, we have then $T_N=0$ at $f_0=1$, which marks the QCP of the local moment antiferromagnetism. Hence the magnetic QCP always coincides with the delocalization QCP, as observed in YbRh$_2$Si$_2$ and CeRhIn$_5$, providing that it is not surrounded by superconductivity or other long-range orders of the itinerant heavy electrons.

\begin{figure}[t]
\centerline{{\includegraphics[width=0.5\textwidth]{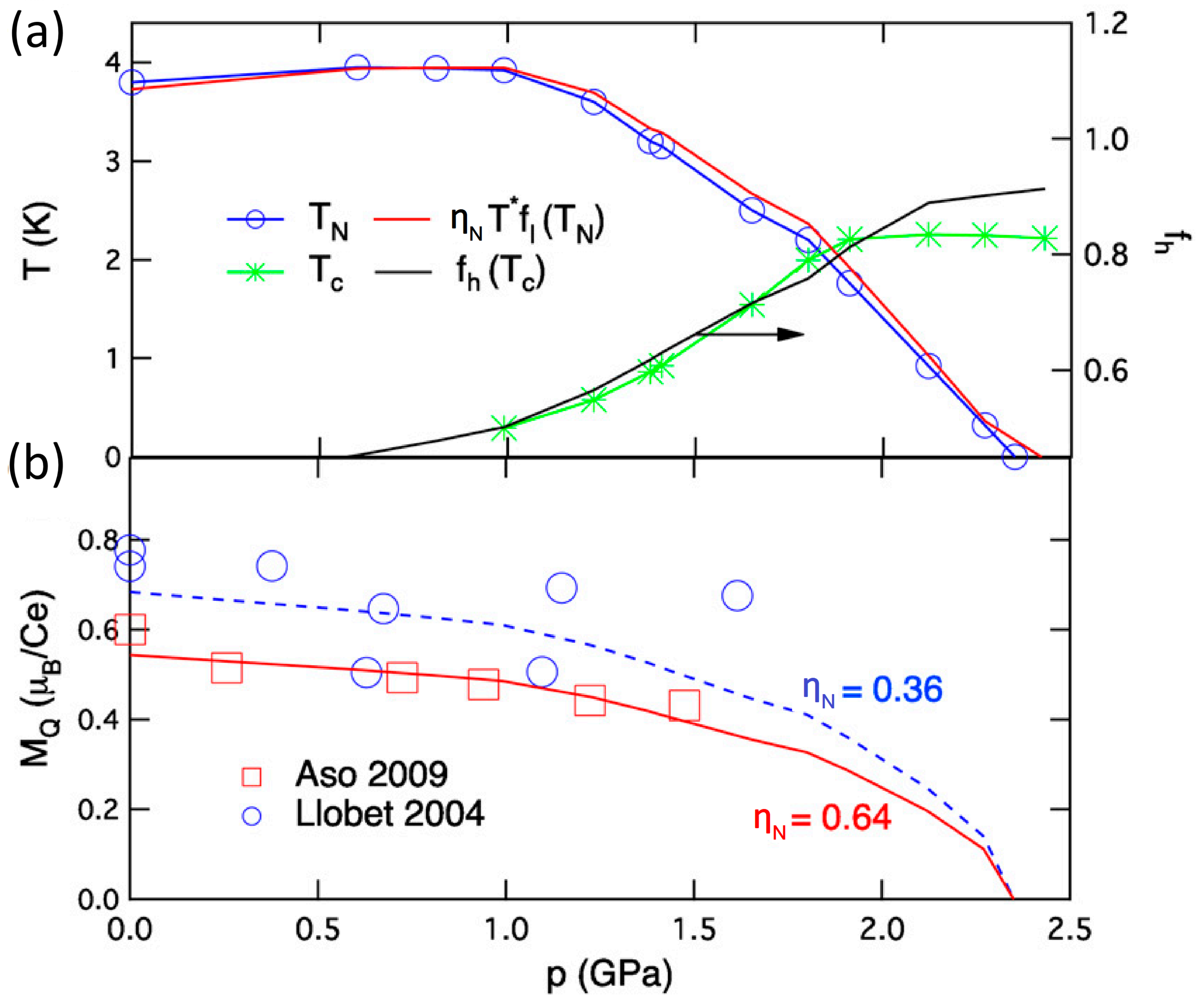}}}
\caption{
{(Color online) Two-fluid fit to the N\'eel temperature and the ordered moment as a function of pressure for CeRhIn$_5$ \cite{Aso2009,Llobet2004}. Different lines indicate the theoretical fit with different values of $\eta_N$. Figure adapted from \cite{Yang2012}.}
\label{Fig13}}
\end{figure}

For the antiferromagnetic state, the magnitude of the ordered moment is approximately given by
\begin{equation}
\mu^2=f_l(T_N)\mu_0^2,
\label{Eq27}
\end{equation}
so that we have the relation,
\begin{equation}
\frac{T_N}{T^*}=\eta_N\frac{\mu^2}{\mu_0^2},
\label{Eq28}
\end{equation}
where $\mu_0$ is the full moment above $T^*$. We may test these formulas for any local moment antiferromagnet if $T^*$ and $f_0$ could be determined from experiment. In general, $T^*(p)$ can be estimated from the coherence temperature in the magnetic resistivity, while the hybridization parameter, $f_0(p)$, cannot be obtained straightforwardly and requires some extra considerations. Detailed analysis for CeRhIn$_5$ and YbRh$_2$Si$_2$ can be found in \cite{Yang2012,Yang2014a}. \Fref{Fig13} shows the fitting results on the N\'eel temperature and the ordered moments in CeRhIn$_5$ \cite{Aso2009,Llobet2004} and the overall agreement with experiment is quite good.

\begin{figure}[t]
\centerline{{\includegraphics[width=0.5\textwidth]{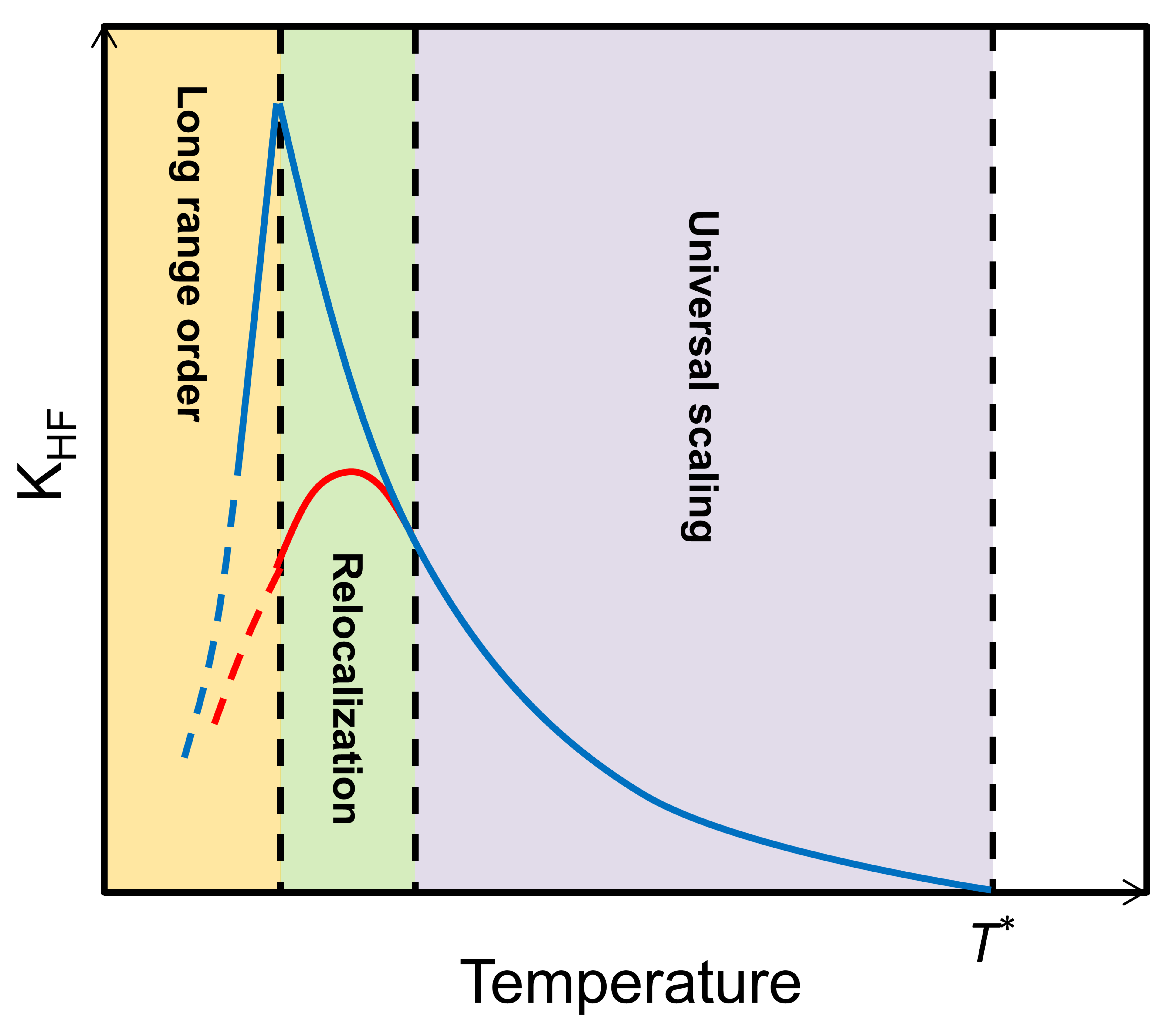}}}
\caption{
{(Color online) Illustration of the temperature evolution of the Knight shift anomaly approaching different low temperature orders. Figure adapted from \cite{Shirer2012}.}
\label{Fig14}}
\end{figure}

We would like to point out a peculiar effect in heavy electron antiferromagnet, namely the relocalization of itinerant heavy electrons in the approach to magnetic ordering. This effect was first observed in the Knight shift anomaly. \Fref{Fig14} summarizes the different situations of the Kondo liquid evolution in response to different long-range orders \cite{Shirer2012}. In contrast to the case of CeCoIn$_5$, CeIrIn$_5$ (superconductor) and URu$_2$Si$_2$ (hidden order), where the Kondo liquid susceptibility increases continuously from above $T^*$ to the ordering temperature without showing any signature of saturation, those in the antiferromagnets CeRhIn$_5$ \cite{Shirer2012} and CePt$_2$In$_7$ \cite{Warren2011} exhibit a maximum and then start to decrease before it reaches $T_N$. The latter reflects the precursor effect to the long-range magnetic order due to the onset of strong antiferromagnetic correlations as observed in the inelastic neutron scattering measurement \cite{Bao2000} and the NMR spin-lattice relaxation \cite{Curro2003}. It indicates a subtle balance between the two fluids and suggests a reverse transfer (relocalization) of the f-electron spectral weight from the heavy electron component to the local moment component as the latter develops long-range antiferromagnetic correlations and eventually gets ordered. The relocalization effect reflects the interaction between two fluids and may help us understand the driving force of the magnetic ground states.

\subsection{Superconductivity}

In most heavy electron materials, unconventional superconductivity arises at the border of antiferromagnetic long-range order and the pairing glues are believed to be associated with the magnetic quantum critical fluctuations. It is, however, difficult to develop a complete theory because of the unusual normal state from which superconductivity emerges. In this section, we provide experimental evidences for the pairing condensation of the Kondo liquid. This leads us to propose a simple phenomenological model for the effective attractive quasiparticle interaction and a BCS-like  formula for the transition temperature, $T_c$.

\subsubsection{The Kondo liquid condensation}

\begin{figure}[t]
\centerline{{\includegraphics[width=0.5\textwidth]{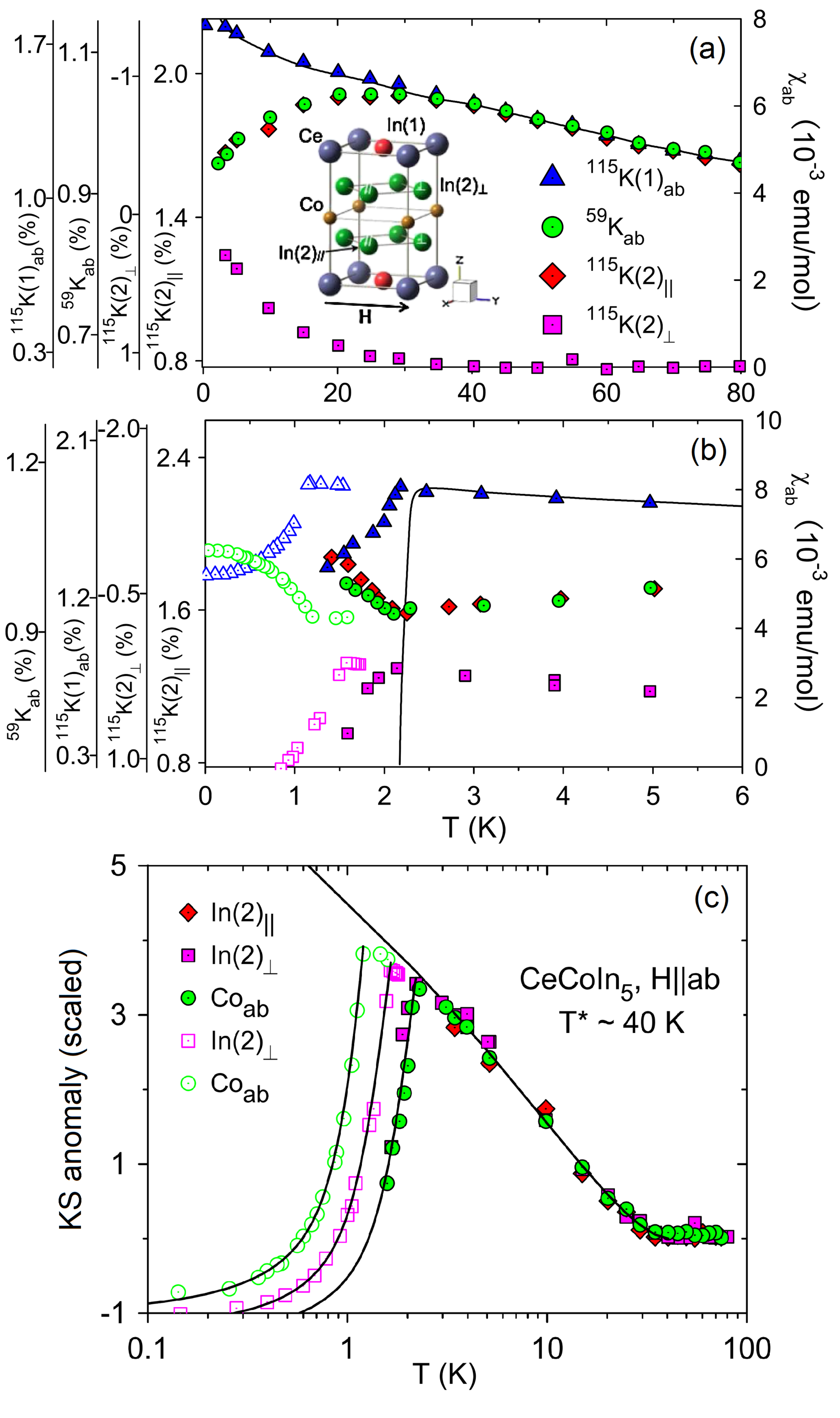}}}
\caption{
{(Color online) (a) Planar Knight shift and magnetic susceptibility of CeCoIn$_5$ above and below $T_c$ \cite{Curro2001}; (b) The subtracted Knight shift anomaly above and below $T_c$ \cite{Yang2009a}. The solid lines are the two-fluid and BCS fit, respectively. Figure adapted from \cite{Yang2009a}.}
\label{Fig15}}
\end{figure}

As discussed earlier, analysis of the spin-lattice relaxation rate in CeCoIn$_5$ indicates that the Kondo liquid may exhibit 2D quantum critical spin fluctuations \cite{Yang2009a}. This provides possible pairing glues for its superconductivity. Direct experimental evidences for the superconducting condensation of the Kondo liquid come from the analysis of the Knight shift anomaly in CeCoIn$_5$. As shown in \fref{Fig15}(a), its planar Knight shift data have two special features \cite{Yang2009a}. First, no anomaly is observed at the In(1) site, indicating a cancellation of the In(1) hyperfine couplings to the two fluids [see \eref{Eq5&6}]. Thus In(1) probes the total spin susceptibility in the whole temperature range, even below $T_c$ where the spin susceptibility cannot be directly measured. Second, the planar Knight shift at the In(2)$_\perp$ site is constant above $T^*$, but becomes temperature dependent below $T^*$. This indicates that In(2)$_\perp$ only probes the Kondo liquid.

These features allow us to use In(1) to subtract the Knight shift anomaly at other sites and use In(2)$_\perp$ as an independent check. The subtracted results at different sites are plotted in \fref{Fig15}(b). We see that they all fall upon the same curve of the Kondo liquid scaling in the normal state, and exhibit similar suppression in the superconducting state, with slightly different $T_c$ due to the difference in the applied magnetic fields. The suppression follows exactly the prediction of the Bardeen-Cooper-Schrieffer (BCS) theory for the d-wave superconductivity \cite{Yang2009a},
\begin{equation}
K_{anom}(T)-K_{anom}(0)\propto\int dE\left(-\frac{\partial f_{FD}(E)}{\partial E}\right)N(E),
\label{Eq29}
\end{equation}
where $f_{FD}(E)$ is the Fermi distribution function, $N(E)\propto\left\langle |E|\sqrt{E^2-\Delta_k(T)^2}\right\rangle_{FS}$ is the average density of states, and $\Delta_k(T)$ is the $\mathbf{k}$-dependent superconducting gap. \Fref{Fig15}(b) gives the best fit of the Knight shift anomaly below $T_c$. We obtain the maximal gap amplitude, $\Delta(0)/T_c\sim 4.5$, which is in good agreement with previous estimates \cite{Sidorov2002,Kohori2006}. This supports the idea that the unconventional superconductivity originates from the unusual normal state, the heavy electron Kondo liquid.

\subsubsection{A spin fluctuation model}

The dominance of superconductivity around the QCP suggests that the coupling of quantum critical spin fluctuations to the heavy electron quasiparticles plays a central role. Insights on the superconducting pairing of the Kondo liquid may be obtained following a microscopic calculation of quantum critical spin-fluctuation induced superconductivity resembling to that for cuprates. The effective pairing interaction may be written as \cite{Monthoux2007}
\begin{equation}
V(\mathbf{q},\omega)=g^2\chi(\mathbf{q},\omega),
\label{Eq30}
\end{equation}
where $g$ is the quasiparticle-spin fluctuation coupling strength and $\chi(\mathbf{q},\omega)$, the dynamic susceptibility, follows the typical Millis-Monien-Pines (MMP) form due to its proximity to an antiferromagnetic state \cite{Millis1990},
\begin{equation}
\chi(\mathbf{q},\omega)=\frac{\chi_\mathbf{Q}}{1+(\mathbf{q}-\mathbf{Q})^2\xi^2-\rmi\omega/\omega_{SF}},
\label{Eq31}
\end{equation}
where $\mathbf{Q}$ is the ordering wave vector, $\omega_{SF}$ is a temperature-dependent spin fluctuation energy, $\chi_\mathbf{Q}=\pi\chi_0(\xi/a)^2$ is the spin susceptibility at $\mathbf{Q}$, $\xi$ is the antiferromagnetic correlation length, $a$ is the lattice constant, and $\chi_0$ is the uniform spin susceptibility. Although a strong coupling calculation has yet to be carried out for heavy electron materials, it is expected to yield a BCS-like expression in analogy to that found for the cuprates \cite{Monthoux1991,Monthoux1992}, namely,
\begin{equation}
T_c=\lambda_1\omega_{SF}(\xi/a)^2\exp\left(-\frac{1}{\lambda_2g\rho_{KL}(T_c)}\right),
\label{Eq32}
\end{equation}
where $\lambda_1$ and $\lambda_2$ are constants of order unity and $\rho_{KL}(T_c)$ is the heavy electron density of states at $T_c$.

\subsubsection{A phenomenological BCS-like formula}

Similar to the conventional BCS formula, the above formula of $T_c$ depends on three quantities: the quasiparticle density of states, $\rho_{KL}(T_c)$, the average strength, $g$, of the induced attractive interaction between quasiparticles, and the average energy range, $\omega_{SF}(\xi/a)^2$, over which it is attractive. Several experimental observations have provided important clues for the determination of these parameters: First, since the Kondo liquid is responsible for the superconductivity, the quasiparticle density of states can be estimated using the Kondo liquid formula \eref{Eq3}; Second, because the Kondo liquid is born out of interacting local moments, its effective quasiparticle interaction is expected to be, $V=\eta T^*$, where $T^*$ is the RKKY interaction between local moments and $\eta$ is a parameter that measures the relative effectiveness of spin fluctuations in bringing about superconductivity for a given material; Third, as first noticed by Pines \cite{Pines2013}, $T_c$ roughly scales with the coherence temperature, $T^*_m$ , at the optimal pressure in quantum critical superconductors, which suggests that $T^*_m$ plays the role of the Debye temperature in the conventional BCS theory and sets the range of energies over which the quantum critical spin-fluctuation induced interaction will be attractive. Combining these observations yields the following BCS-like formula \cite{Yang2014b},
\begin{equation}
T_c(p)=0.14T^*_m\exp\left(-\frac{1}{\rho_{KL}(p,T_c)V(p)}\right)=0.14T^*_m\exp\left(-\frac{1}{\eta\kappa(p)}\right),
\label{Eq33}
\end{equation}
where we have introduced the dimensionless coupling, $\kappa(p)=\rho_{KL}(p,T_c)T^*(p)$. The logarithmic divergence in the density of states of the Kondo liquid and hence $\kappa(p)$ are cut off at low temperatures due to either complete delocalization at $T_L$, or long-range magnetic orders for $f_0<1$, or superconductivity itself at $T_c$. We have \cite{Yang2014b}
\begin{equation}
\kappa(p)=\frac{3\ln2}{2\pi^2}f_0(p)\left(1-\frac{T_{cutoff}(p)}{T^*(p)}\right)^{3/2}\left(1+\ln\frac{T^*(p)}{T_{cutoff}(p)}\right),
\end{equation}
where $T_{cutoff}=T_L$, $T_c$, or $T_{0/N}$, depending on the low temperature orders. 

\begin{figure}[t]
\centerline{{\includegraphics[width=0.8\textwidth]{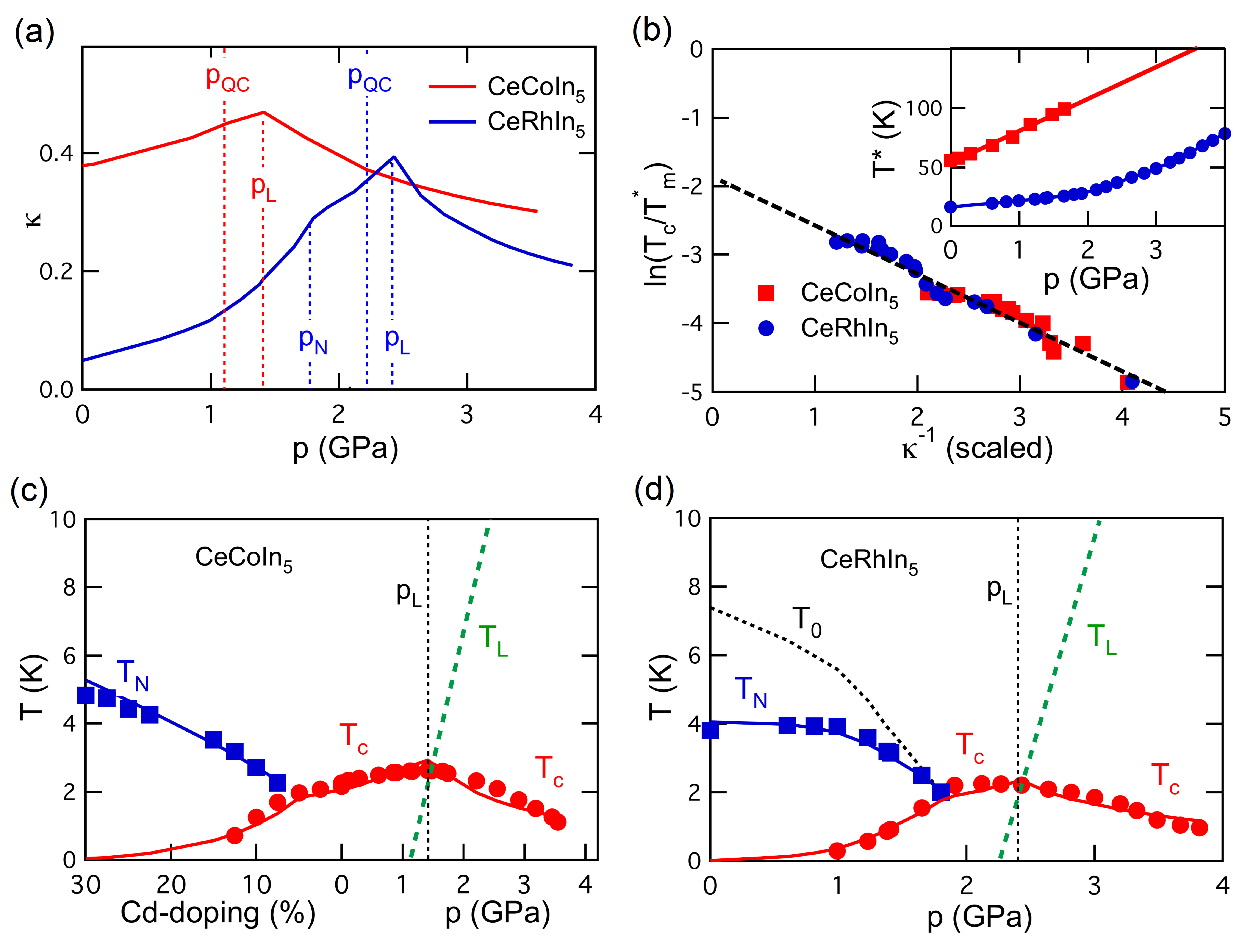}}}
\caption{
{(Color online) Comparison between theory and experiment for the superconductivity in CeCoIn$_5$ \cite{Nicklas2001,Pham2006} and CeRhIn$_5$ \cite{Park2006}. (a) The dimensionless coupling $\kappa(p)$ as a function of pressure; (b) Linear relation between $\ln(T_c/T^*_m)$ and $\kappa(p)^{-1}$; (c) and (d) Fit to the N\'eel temperature and the superconducting transition temperature in CeCoIn$_5$ and CeRhIn$_5$. Figure adapted from \cite{Yang2014b}.}
\label{Fig16}}
\end{figure}

Using experimental data for $T^*(p)$ (the coherence temperature) and the cutoff temperature and assuming that $f_0(p)$ varies linearly with $T^*(p)$, we can estimate the value of $\kappa(p)$ (see the appendix in \cite{Yang2014b} for more details). Figures \ref{Fig16}(a) and \ref{Fig16}(b) show the pressure dependence of $\kappa(p)$ and the comparison between $1/\kappa(p)$ and $\ln(T_c/T^*_m)$ for both CeCoIn$_5$ and CeRhIn$_5$. The good linearity confirms the validity of our BCS-like equation, with a common intercept that leads to the prefactor $0.14T^*_m$ in the above formula. Neutron scattering measurements of the spin fluctuation spectra near $T_c$ at ambient pressure yield $\omega_{SF}=0.3\pm0.15\,$meV and $\xi=9.6\pm1.0\,$\AA (about twice the in-plane lattice constant $a=4.60\,$\AA) in CeCoIn$_5$. We have $\omega_{SF}(\xi/a)^2=1.3\,$meV$\sim15.1\,$K, in close agreement with above phenomenological result, $0.14T^*_m=12.9\,$K. Figures \ref{Fig16}(c) and \ref{Fig16}(d) show our fit to the experimental data with $\eta=1.30$ for CeCoIn$_5$ \cite{Nicklas2001,Pham2006} and $\eta=3.09$ for CeRhIn$_5$ \cite{Yang2014b}. The dome structure of the superconducting $T_c$ is well explained, as well as the pressure and doping variation of $T_N$, both in remarkably good agreement with experiment. 

\subsubsection{A generic phase diagram}

\begin{figure}[t]
\centerline{{\includegraphics[width=0.5\textwidth]{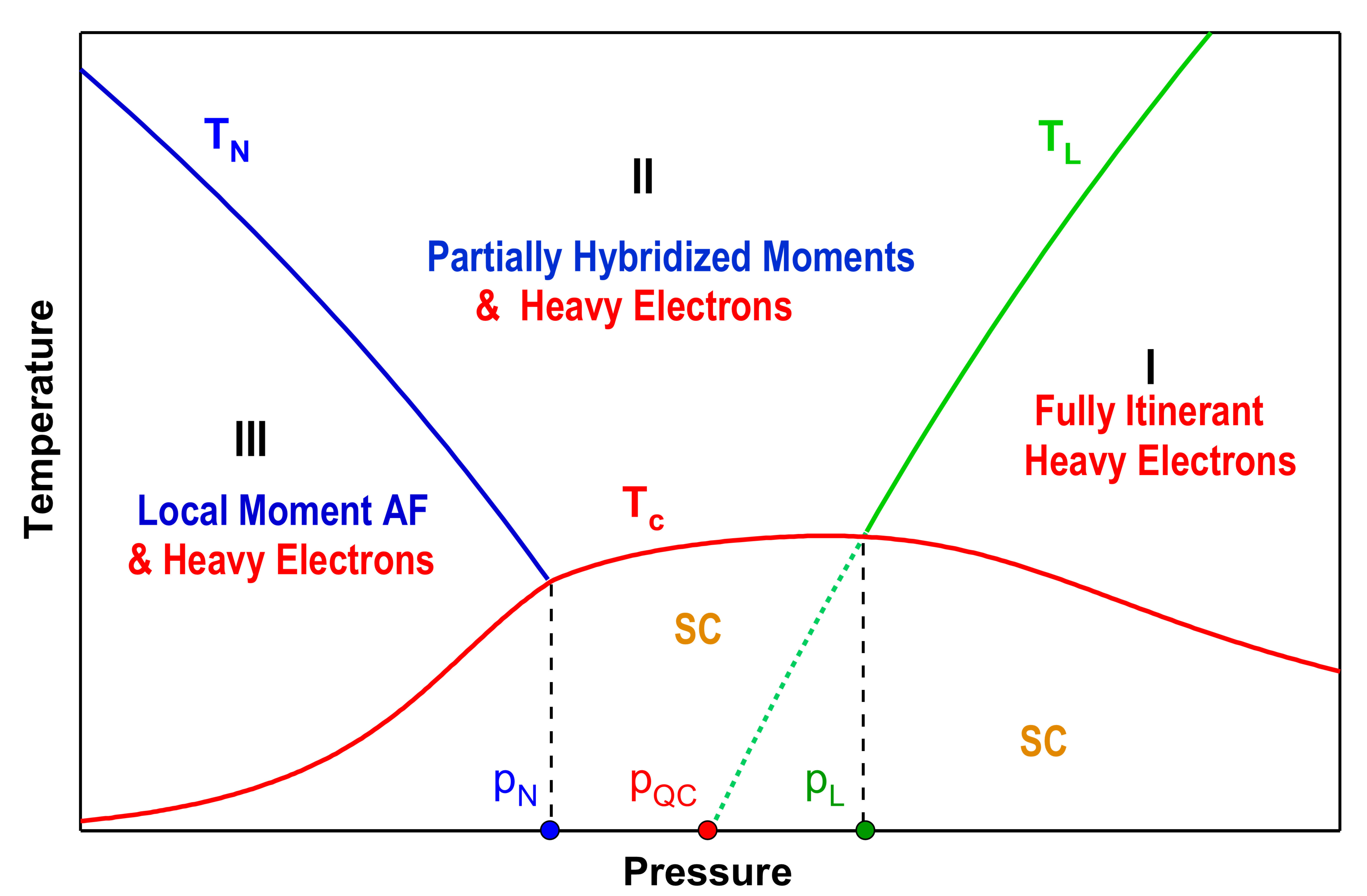}}}
\caption{
{(Color online) Illustration of the superconducting phase diagram in the two-fluid model. Three regions are identified where superconductivity emerges out of fully itinerant heavy electrons (I), coexists with local moment antiferromagnet (III) or coexists with residual unhybridized and disordered moments (II). Figure adapted from \cite{Yang2014b}.}
\label{Fig17}}
\end{figure}

Our BCS-like formula leads to a generic phase diagram of heavy electron quantum critical superconductors, as shown in \fref{Fig17}. Taking into account the low temperature cutoff by the various ordered states, we can identify three regimes of superconductivity \cite{Yang2014b}:
\begin{itemize}
\item Region I: $T_c\le T_L$. Superconductivity emerges from a fully formed heavy electron state. The increase in $T_c$ with decreasing pressure is brought about by the enhancement in the heavy electron density of states produced by the decrease in $T_L$, so $T_c$ reaches its maximal value at the pressure, $p_L$, at which the superconducting transition and the delocalization line intersect. Since $T_{cutoff}(p_L)=T_c(p_L)=T_L(p_L)$ and $f(T_c,p_L)=1$, we have $\kappa(p_L)=3\ln2/2\pi^2\left[1+\ln(T^*_m/T^{max}_c)\right]$. The value of  $T^{max}_c/T^*_m$ depends only on the value of $\eta$, the impedance match between the spin-fluctuation spectra and the heavy electron Fermi surface. Following Monthoux and Lonzarich \cite{Monthoux2001,Monthoux2002}, these variations can be understood from the change in the effective dimensionality and the crystal structure in each material.
\item Region II: $T_c>T_L$ and $T_N$. Superconductivity emerges from a partially formed heavy electron state whose ability to superconduct is reduced by the residual unhybridized local moments with which it coexists. The QCP is located in this region and provides the paring glues for all three regions. It is interesting to see if there may still exist unhybridized local moments deep inside the superconducting phase.
\item Region III: $T_c\le T_N$. The residual unhybridized local moments get ordered at the N\'eel temperature $T_N$, coexisting with the remnant heavy electrons that become superconducting at lower temperatures. The decrease in $T_c$ with decreasing pressure arises from the reduction in the heavy electron density of states brought about by the partial relocalization of the heavy electrons.
\end{itemize}
The proposed phase diagram is consistent with experimental observations and provides a natural explanation to the dome structure of heavy electron quantum critical superconductors such as CeRhIn$_5$.

\subsection{Quantum criticality}

Quantum criticality plays an important role in heavy electron materials. It leads to anomalous scaling properties in the normal state and provides pairing glues for heavy electron superconductivity. Recently, it was found that the QCP can be tuned by an external magnetic field, giving rise to the field-induced quantum criticality, as observed in YbRh$_2$Si$_2$ \cite{Ishida2002}, or a quantum critical line on the pressure-magnetic field phase diagram, as observed in CeCoIn$_5$ \cite{Zaum2011,Ronning2006}. One may wonder whether the two-fluid model could explain such behaviors. In this section, we discuss how magnetic field may interplay with the two-fluid physics.

\subsubsection{Field induced change in the hybridization effectiveness}

In the two-fluid model, the quantum critical point is the end point of the delocalization line at $T_L=0$. As discussed previously, the delocalization line is determined by $f(T_L,p,H)=1$, marking a crossover from partially localized to fully itinerant behavior of the f-electrons. To get $T_L=0$ requires $f_0(p,H)=1$. Hence to study how magnetic field may tune the QCP, we need to consider its influence on $f_0$, which to the lowest order approximation may be written as
\begin{equation}
f_0(p,H)=f_0(p)\left[1+\left(\eta_H H\right)^\alpha\right],
\label{Eq34}
\end{equation}
where $\alpha$ is a scaling parameter. In the vicinity of the quantum critical point, we may also expand $f_0(p)$ as
\begin{equation}
f_0(p)\approx 1+\eta_p\left(p-p^0_c\right),
\label{Eq35}
\end{equation}
where $p^0_c$ is the quantum critical pressure at $H=0$. For simplicity, we assume $\alpha$, $\eta_p$ and $\eta_H$ are all field-independent constant and explore in the following the consequences of above approximations. For Ce-compounds, collective hybridization is enhanced with increasing pressure so $\eta_p>0$, whereas for Yb-compounds, collective hybridization is suppressed with increasing pressure and $\eta_p<0$. For both compounds, we assume that local hybridization is enhanced by the magnetic field.

\subsubsection{Quantum critical and delocalization lines}

At zero temperature, $f_0(p,H)=1$ predicts a line of quantum critical points on the pressure-magnetic field plane \cite{Yang2014a}. We have
\begin{equation}
p_c(H)=p^0_c-\frac{1}{\eta_p}\frac{\eta_H^\alpha H^\alpha}{1+\eta_H^\alpha H^\alpha}.
\label{Eq36}
\end{equation}
At ambient pressure, the delocalization temperature can also be obtained as, 
\begin{equation}
\frac{T_L(H)}{T^*}=1-\left(\frac{1+\eta_H^\alpha H_{QC}^\alpha}{1+\eta_H^\alpha H^\alpha}\right)^{2/3},
\label{Eq37}
\end{equation}
where $H_{QC}$ is the critical field at ambient pressure. Both the quantum critical line and the delocalization line are determined by the same scaling parameter, $\alpha$.

\begin{figure}[t]
\centerline{{\includegraphics[width=0.5\textwidth]{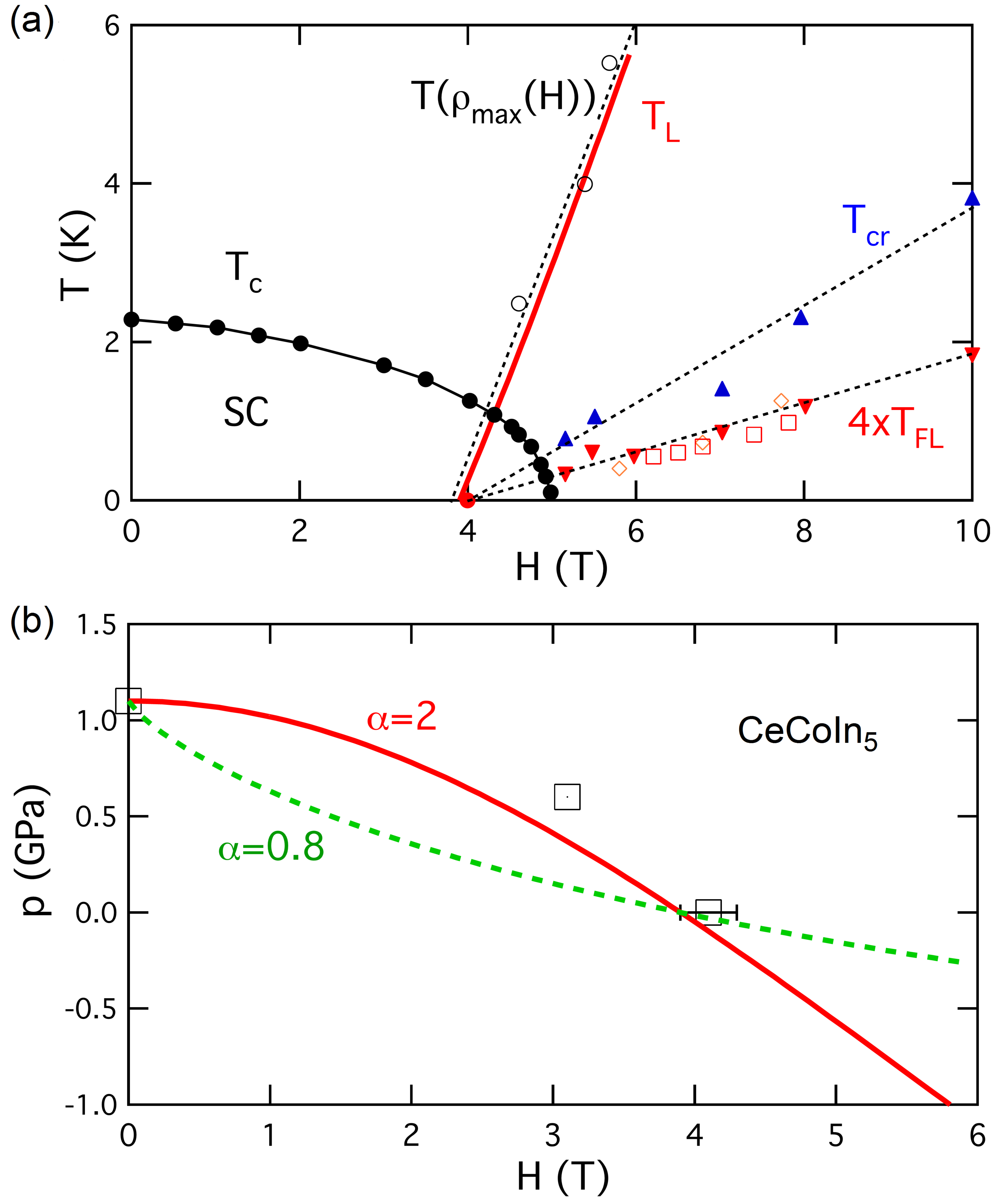}}}
\caption{
{(Color online) Two-fluid fit with $\alpha=2$ for CeCoIn$_5$ \cite{Zaum2011,Ronning2006,Paglione2003,Singh2007}. (a) The delocalization temperature as a function of the magnetic field; (b) The quantum critical line on the pressure-field phase diagram. Figure adapted from \cite{Yang2014a}.}
\label{Fig18}}
\end{figure}

The above results have been tested in CeCoIn$_5$ and YbRh$_2$Si$_2$ \cite{Yang2014a}. In CeCoIn$_5$, a joint analysis of resistivity and thermal expansion data has led Zaum \etal \cite{Zaum2011} to determine the quantum critical field $H_{QC}=4.1\pm0.2\,$T inside the superconducting dome at ambient pressure. We note that the exact location of $H_{QC}$ is still under debate and some suggest a zero-field quantum critical point in CeCoIn$_5$ \cite{Tokiwa2013}. Nevertheless, several temperature scales have been identified in the $H-T$ phase diagram as shown in \fref{Fig18}(a) \cite{Zaum2011,Ronning2006,Paglione2003,Singh2007}. On the other hand, scaling analysis of the magneto-resistivity suggests a quantum critical pressure, $p^0_c=1.1\,$GPa, at zero magnetic field \cite{Ronning2006}. This difference leads to the idea of a quantum critical line in the $p-H$ plane as shown in \fref{Fig18}(b). Similar results have also been investigated in YbRh$_2$Si$_2$ \cite{Ishida2002,Gegenwart2002,Knebel2005,Tokiwa2009,Pfau2012}, which has a N\'eel temperature of 0.07 K at ambient pressure. The antiferromagnetic order is suppressed with a critical field, $H_{QC}=0.055\,$T, along the easy-axis. At high field, a characteristic temperature scale has been observed in many measurements and found to coincide with the magnetic quantum critical point at zero temperature. It is thus identified as the delocalization line in the two-fluid model. 

\begin{figure}[t]
\centerline{{\includegraphics[width=0.5\textwidth]{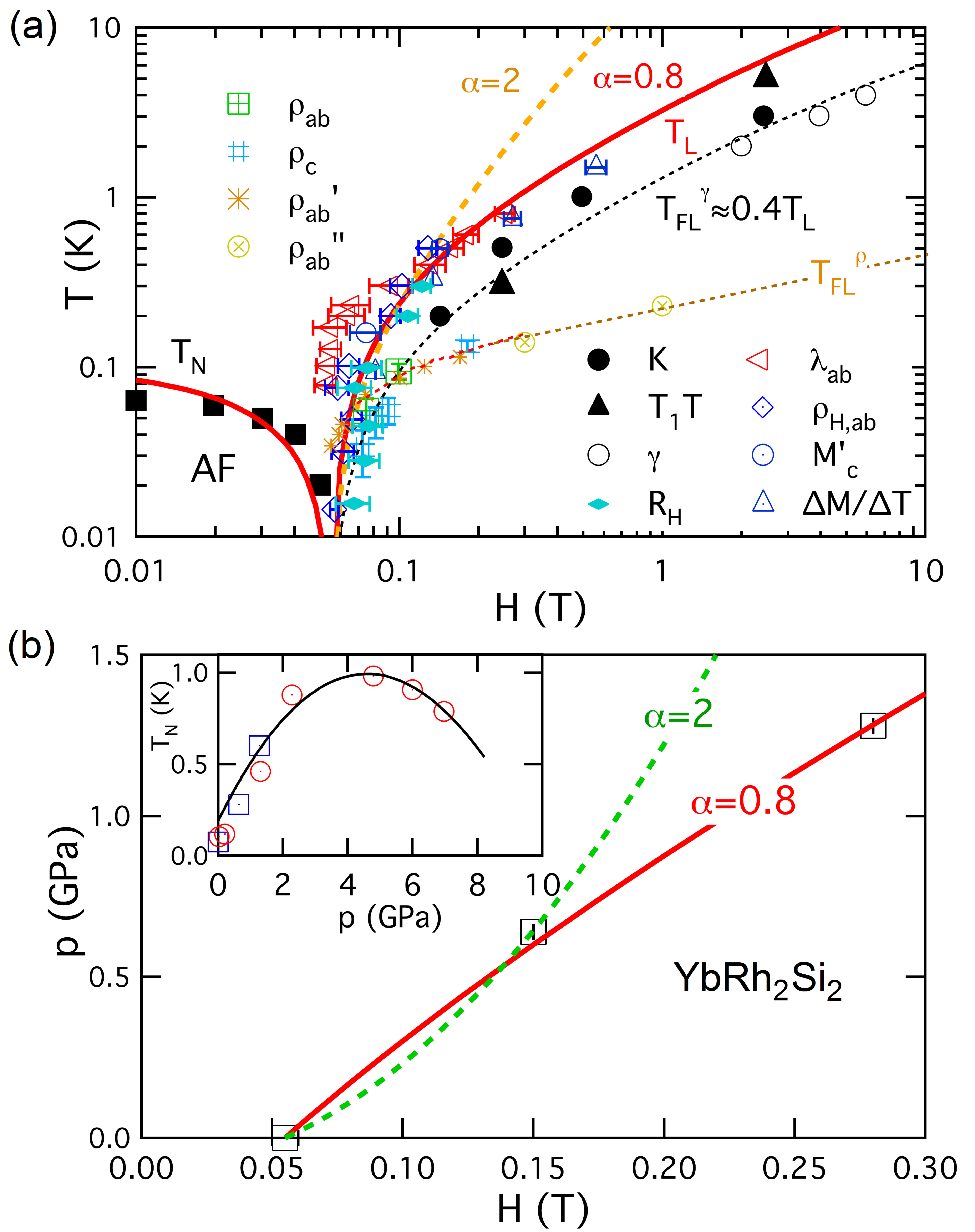}}}
\caption{
{(Color online) Two-fluid fit with $\alpha=0.8$ for YbRh$_2$Si$_2$ \cite{Ishida2002,Gegenwart2002,Knebel2005,Tokiwa2009,Pfau2012}. (a) The N\'eel temperature and the delocalization temperature as a function of the magnetic field; (b) The quantum critical line on the pressure-field phase diagram. The inset shows the fit to the pressure variation of the N\'eel temperature. Figure adapted from \cite{Yang2014a}.}
\label{Fig19}}
\end{figure}

Figures \ref{Fig18} and \ref{Fig19} show the two-fluid fit to the delocalization line and the quantum critical line in CeCoIn$_5$ and YbRh$_2$Si$_2$. The good agreement confirms once again the two-fluid prediction. More detailed analysis can be found in \cite{Yang2014a} and will not be repeated here. We only note that the very different quantum critical behaviors of the two compounds seem to be fully incorporated in their different values of the scaling parameter: $\alpha=2$ for CeCoIn$_5$ and $\alpha=0.8$ for YbRh$_2$Si$_2$. 

\subsubsection{Quantum critical scaling}

Quantum critical scaling in other quantities of interest can be readily obtained if we take $T_L(H)$ as the fundamental energy scale of the Fermi liquid state. Assuming a power-law scaling in the vicinity of the quantum critical point, we obtain a simple expression for the effective mass \cite{Yang2014a}, 
\begin{equation}
\frac{m^*(H)}{m_0}=\left(\frac{T^*}{T_L(H)}\right)^{\alpha/2},
\label{Eq38}
\end{equation}
in which the scaling exponent, $\alpha/2$, is chosen based on experimental analysis, and $m_0$ is the bare mass. The specific heat coefficient is then given by
\begin{equation}
\gamma_{QC}(H)=\gamma_0\left(\frac{T^*}{T_L(H)}\right)^{\alpha/2},
\label{Eq39}
\end{equation}
where $\gamma_0$ is independent of the magnetic field. This formula is different from the Kondo liquid scaling in \eref{Eq25}, reflecting the influence of quantum criticality. If we further assume a constant Kadowaki-Woods ratio, $A(H)/\gamma(H)^2$, where $A(H)$ is the resistivity coefficient defined in $\rho(T,H)\sim A(H)T^2$, we obtain immediately a third scaling formula,
\begin{equation}
A(H)=\frac{A_0}{T^*{}^2}\left(\frac{T^*}{T_L(H)}\right)^{\alpha},
\label{Eq40}
\end{equation}
where $A_0$ is a field-independent prefactor.
\begin{figure}[t]
\centerline{{\includegraphics[width=0.5\textwidth]{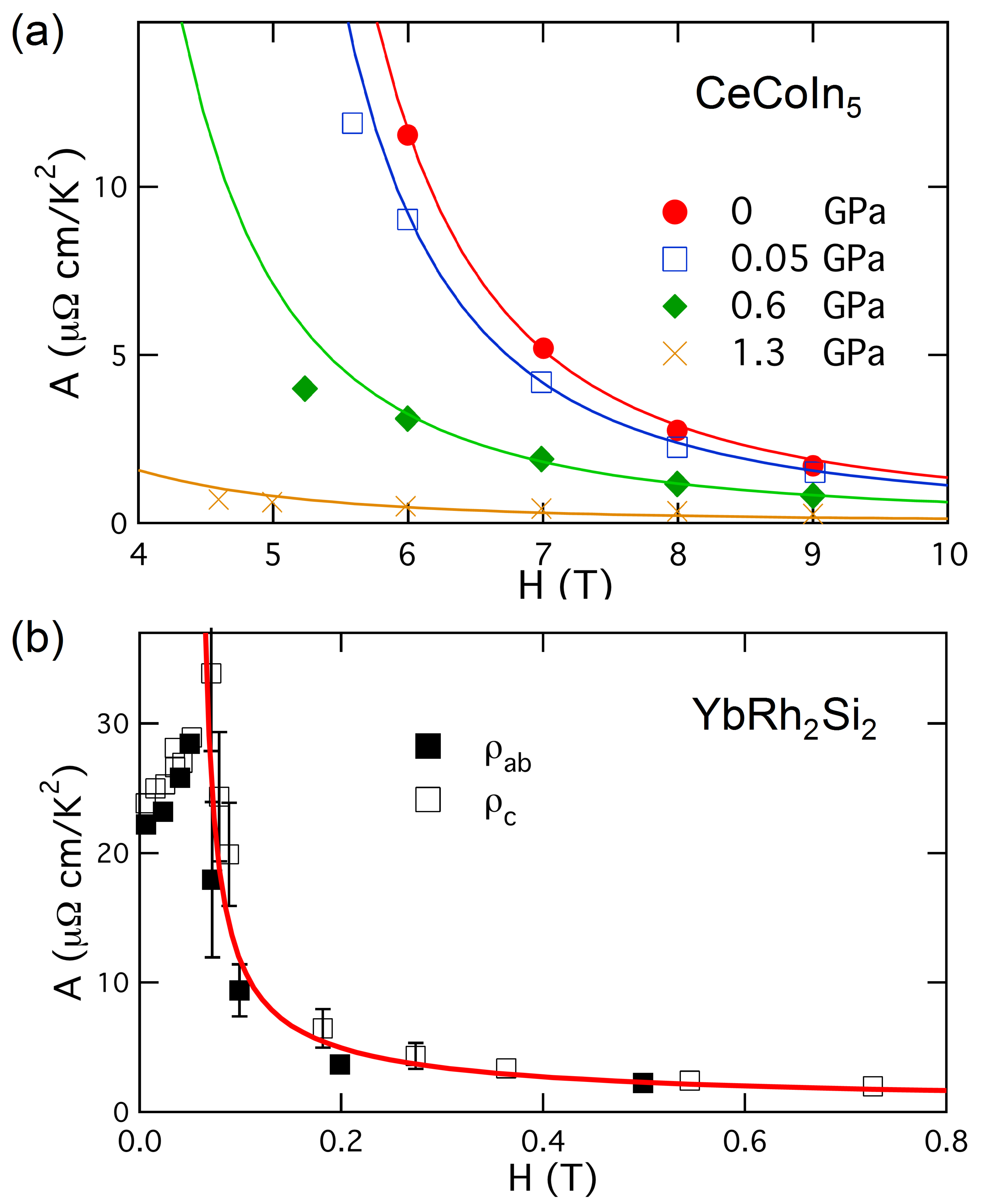}}}
\caption{
{(Color online) Two-fluid fit to the resistivity coefficient in (a) CeCoIn$_5$ \cite{Ronning2006} and (b) YbRh$_2$Si$_2$ \cite{Gegenwart2002}. Figure adapted from \cite{Yang2014a}.}
\label{Fig20}}
\end{figure}

To test the above results, \fref{Fig20} plots the field dependence of the resistivity coefficient in CeCoIn$_5$ and YbRh$_2$Si$_2$ \cite{Ronning2006,Gegenwart2002}. We find that $\alpha=2$ and 0.8 yield good fit to the experimental data in the two compounds, respectively. The fact that this same parameter can be used to explain several seemingly unrelated physical properties implies the predictive power of the two-fluid model. The nature of the scaling parameter and what determines its exact value are subject to future studies.

\section{Concluding remarks}

Different from the single-ion Kondo problem \cite{Hewson1993}, the Kondo lattice problem is still unsolved and under intensive debate. We have shown that the two-fluid model can explain a large variety of experimental data that cover the magnetic, electronic, transport and thermal properties of many heavy electron materials. It therefore provides a simple and unified framework for understanding the heavy electron physics. However, we should note that the underlying mechanism of the two-fluid behavior is still unclear. Especially, the universal scaling that is predicted in the two-fluid model and examined in the Knight shift, the spin-lattice relaxation rate and the Hall coefficient has not been explained in any current theory and needs particular attention in future investigations \cite{Barzykin2006,Shim2007,Zhu2011,Choi2012,Jiang2014,Xie2015}. We also see that $T^*$ is typically larger than the single ion Kondo temperature and the fact that it is given by the RKKY interaction is distinctly different from the conventional way of thinking that $T^*$ originates from the Kondo temperature. Our study on the one-dimensional Kondo-Heisenberg model using the exact density matrix renormalization group (DMRG) method suggests that the two-fluid behavior results from the simultaneous spin entanglement of the local moments with one another and the conduction electrons. We find that antiferromagnetic spin fluctuations do not always kill the heavy electrons and can actually enhance collective hybridization in some parameter range \cite{Xie2015}. We expect that similar physics should work in realistic materials. More discussions on the implications of the two-fluid model on the microscopic theory can be found in \cite{Lonzarich2016}.

Two future experiments may be crucial for achieving a better understanding of the underlying mechanism. One is the measurement of the Fermi surface change with temperature, which has so far not been thoroughly investigated due to technical limitations. It will provide a further examination of the two-fluid prediction and may help to establish detailed understanding of the unusual electronic structures of the heavy electrons in the momentum space and reveal the basic mechanism governing the heavy electron emergence. The other is the direct detection of the two fluids. A recent experiment has observed two different components near the quantum critical point in YbRh$_2$Si$_2$ \cite{Kambe2014}. Detection of the two coexisting fluids using ultrafast or other techniques may provide a decisive justification of the two-fluid physics.

The Kondo lattice materials are in many ways the simplest correlated electron materials, where charge fluctuations of the f-electrons are suppressed. Similar two-fluid behavior has also been observed in cuprates and iron-based compounds \cite{Barzykin2009,Dai2012,Ji2013}. One may therefore speculate that the two-fluid physics is a generic feature of correlated electrons that locate at the border of localization and itinerancy. Our study of the heavy electron physics may provide the key for understanding the physics of all strongly correlated electron systems. 

\ack
We thank D. Pines, Z. Fisk, J. D. Thompson, N. J. Curro and G. Lonzarich for the many stimulating discussions. This work was supported by the State Key Development Program for Basic Research of China (Grant No. 2015CB921303), the National Natural Science Foundation of China (NSFC Grant No. 11522435) and the Strategic Priority Research Program of the Chinese Academy of Sciences (Grant No. XDB07020200). We thank the Simons Foundation for its support and the Aspen Center for Physics (NSF Grant PHY-1066293) for the hospitality during the writing of this paper.

\end{document}